\documentclass{elsart}

\usepackage{amsmath,amsfonts,latexsym}
\usepackage[square]{natbib}
\usepackage{graphicx}
\usepackage{dcolumn}
\usepackage{bm}
\newcommand{\calL}{{\mathcal L}}
\newcommand{\calG}{{\mathcal G}}
\newcommand{\calO}{{\mathcal O}}
\newcommand{\calD}{{\mathcal D}}
\def\comment#1{}

 \newcommand{\BF}[1]{\mbox{\boldmath $#1$}}
 \newcommand{\BFS}[1]{\mbox{\scriptsize\boldmath $#1$}}
\def\nablab{{\BF \nabla}}
\def\nablabs{{\BFS \nabla}}

\def\comment#1{}

\newcommand{\nc}{\newcommand}
\nc{\beq}{\begin{eqnarray}}
\nc{\eeq}{\end{eqnarray}}
\nc{\scs}{\scriptstyle}
\nc{\setval}{\fmfset{wiggly_len}{3mm} \fmfset{arrow_len}{1.5mm}
    \fmfset{arrow_ang}{13} \fmfset{dash_len}{1.5mm}\fmfpen{0.125mm}
    \fmfset{dot_size}{2thick}}

\begin{document}
\runauthor{Kleinert, Nogueira and Sudb{\o}}
\begin{frontmatter}
\title{Kosterlitz-Thouless-like deconfinement mechanism in the $2+1$
dimensional Abelian Higgs model}
\author[Berlin]{Hagen Kleinert}
\ead{kleinert@physik.fu-berlin.de}
\author[Berlin]{Flavio S. Nogueira}
\ead{nogueira@physik.fu-berlin.de}
\author[Trondheim]{Asle Sudb{\o}}
\ead{asle.sudbo@phys.ntnu.no}


\address[Berlin]{Institut f\"ur Theoretische Physik,
Freie Universit\"at Berlin, Arnimallee 14, D-14195 Berlin, Germany}
\address[Trondheim]{Department of Physics, Norwegian University of
Science and Technology, N-7491 Trondheim, Norway}
\vspace{-1cm}
\begin{abstract}
We point out that the permanent
confinement in a compact $2+1$-dimensional $U(1)$ Abelian Higgs model
is destroyed by
 matter fields in the fundamental representation.
The deconfinement transition is
 Kosterlitz-Thouless like.
The dual theory  is shown to
describe a three-dimensional
gas of point charges with {\it logarithmic} interactions
which arises from an anomalous dimension
of the gauge field caused by
 critical matter field
fluctuations.
The theory
is equivalent
to a
 sine-Gordon-like theory in $2+1$ dimensions
with an {\em anomalous gradient energy\/} proportional to $k^3$.
The Callan-Symanzik equation is used  to demonstrate that this
theory  has a massless and a massive phase. The
renormalization group equations for the fugacity $y(l)$ and stiffness
parameter $K(l)$ of the theory show that
the  renormalization of $K(l)$
 induces an anomalous scaling dimension $\eta_y$ of $y(l)$.
The stiffness parameter of the theory  has a universal jump
at the transition  determined  by  the dimensionality and $\eta_y$.
As a byproduct
of our analysis, we relate the critical coupling of the sine-Gordon-like theory
to an {\it a priori} arbitrary constant that enters into the computation of
critical exponents in the Abelian Higgs model at the charged infrared-stable
fixed point of the theory, enabling a determination of this parameter. This
facilitates the computation of the critical exponent $\nu$  at the charged
fixed point in excellent agreement with one-loop renormalization  group
calculations for the three-dimensional $XY$-model, thus confirming  expectations
based on duality transformations.
\end{abstract}
\end{frontmatter}
\section{Introduction}

Gauge theories in $d=2+1$ dimensions are often
considered as effective theories of  strongly correlated systems in two
spatial dimensions at zero temperature \cite{Affleck,Baskaran,Hsu}. Prominent
examples of systems to which such theories are hoped to be applicable are the
high-$T_c$ cuprates in the underdoped or undoped regime. In the undoped
regime
it is known that spinor QED$_3$
is an effective low energy theory for the
quantum Heisenberg antiferromagnet (QHA)
\cite{Affleck}. It is hoped that one effectively
can account for doping by coupling the gauge theory to a scalar boson representing
the holon part (charge part) of  composite Hubbard-operators describing
{\it projected} electrons, which however do not satisfy simple fermion
commutation relations.
Similar effective theories
 have a long history as useful toy-models in high-energy
physics \cite{ES,FradShe,Chernodub1}, and have recently been suggested
to describe neural networks \cite{Matsui1}.

Of particular
interest in the physics of strongly correlated systems is the
compact version of the $2+1$-dimensional
Abelian Higgs model  with matter fields in the fundamental representation.
This is the model we shall be concerned with in this paper
and for which we shall find the results
summarized in the Abstract.

\subsection{Preliminary remarks}

Our starting point is  the following
Abelian euclidean 
field theory of a scalar matter field
coupled to a massless gauge field
\begin{equation}
\label{GL}
{\calL}_{b}=|(\partial_\mu-iA_\mu^0)\phi_0|^2
+m_0^2|\phi_0|^2+\frac{u_0}{2}|\phi_0|^4,
\end{equation}
where the subscript zero denotes bare quantities.
It corresponds to a theory with a
Maxwell term
\begin{equation}
\label{Maxwell}
{\calL}_M=\frac{1}{4e_0^2}{F_{\mu\nu}^0}^2,
\end{equation}
where $F_{\mu\nu}^0=\partial_\mu A_\nu^0-\partial_\nu A_\mu^0$,
 in which the
 gauge coupling $e_0$ goes to infinity.
This limit
implies the constraint
$j_b^\mu=0$, where  $j_b^\mu=\displaystyle\phi_0^*
\mathop{\partial}^{\leftrightarrow  }{}^\mu\phi_0 $ is the
 boson current.

When deriving {\it effective} theories for the $t-J$ model
we arrive naturally at a {\em compact\/} $U(1)$
lattice gauge field
\cite{Baskaran}.
For  QHA, the gauge symmetry is larger and given by
the gauge group $SU(2)$ \cite{Hsu}. However, in this case a reduced
$U(1)$
formulation is also possible
\cite{Affleck}. Since this
  $U(1)$ is a
subgroup of $SU(2)$, which is a compact group, the $U(1)$
gauge theory of QHA is necessarily a compact Abelian
gauge theory.

It is well known
that a compact $U(1)$ theory of the pure Maxwell type in three dimensions
 confines electric
charges permanently \cite{Polyakov}.
In the literature \cite{Nayak} it is also
 argued that
this permanent confinement
should  be present
if
an additional fermionic field $\psi$ coupled
to the gauge field  by a Lagrangian
\begin{equation}
\label{L}
{\calL}_{f}=\sum_{i=1}^N \bar{\psi}_i(\partial_\mu-iA_\mu^0)\psi_i.
\end{equation}
 This means that the particles
represented by the fields $\psi$ and
$\phi_0$ never have an independent dynamics. In the context of many-body theory,
the Dirac fermion $\psi$ could represent a {\it spinon}, while
$\phi$ represents a {\it holon}. If electric test charges were
permanently confined in the model, then the spinon
and the holon would only appear as composite particles.
In this case it would be impossible to
fractionalize the electron, i.e. spin and charge would always remain
attached to each other.
Spin-charge separation is known to occur in $1+1$
dimensions \cite{Mudry}. There  fermions can be transmuted
into bosons via the so called Jordan-Wigner transformation.
In $2+1$ dimensions the situation is less clear, but for
matter fields in the fundamental representation there is
one circumstance where spin-charge separation is known rigorously
to occur, namely the chiral spin liquid state \cite{Wen}.
However,
the statistics of particles can be changed
as in $1+1$ dimensions.
In the chiral spin liquid, spinons have anyonic statistics
described by a Chern-Simons term \cite{Jackiw}
in the effective gauge theory,
which reflects the breaking of parity and time reversal symmetry.

The lack of consensus about spin-charge separation in  $2+1$-dimensional
compact $U(1)$ matter-coupled gauge theories with matter fields
in the fundamental representation initiated
investigations of other
gauge theories for strongly correlated electron systems.
One of the most promising candidates is
a $Z_2$ gauge field coupled to matter fields \cite{Senthil}.
Similar ideas leading
to electron fractionalization
had  earlier been presented in the condensed matter literature
 \cite{RS,Wen1}.
In $2+1$ dimensions the $Z_2$ theory has a
deconfinement transition \cite{FradShe}. Thus, $Z_2$ gauge theories are
potentially good candidates for describing spin-charge separation without
breaking parity and time reversal symmetries.

The confinement properties of $U(1)$ gauge theories for the cuprates and
the relation to spin-charge separation were recently discussed
from various points of view
\cite{Nayak,Senthil1,Nagaosa,Matsui,Nayak1}. Nayak \cite{Nayak}
states that in gauge theories of the $t-J$ model fermions and bosons interact
at infinite (bare) gauge coupling and, for this reason, it is necessarily
a theory with permanent confinement of slave particles.
In contrast, Ichinose and Matsui \cite{Matsui}
have argued that the coupling
to matter fields strongly influences the phase structure of the system. In
Ref. \cite{Nayak1}, it is correctly pointed out  that if spin-charge separation
occurs, it is not necessaryly tied to the notion of 
confinement-deconfinement of slave particles. 
The picture proposed in Ref. \cite{Nayak}
in $2+1$ dimensions is reminiscent of $1+1$ dimensions where
spinons and holons are solitons and cannot be identified with the slave
particles, which are not part of the spectrum \cite{Mudry}.
Nagaosa and Lee \cite{Nagaosa} discuss a compact $U(1)$ gauge
theory coupled to bosonic matter field in the fundamental representation.
They conclude that in $d=2+1$ this theory permanently confines
electric charges, in contrast to the analysis by Einhorn and Savit on the
same model \cite{ES}.

In a recent letter \cite{KNS}, we have studied
 the confining properties
of the Lagrangian (\ref{GL}), as well as the case of a fermionic field
$\psi$  coupled to a gauge field, but with an added Maxwell
term. The Lagrangian (\ref{GL}) with a Maxwell term  corresponds essentially
to the model considered by Nagaosa and Lee \cite{Nagaosa}, though these
authors have considered a frozen-amplitude version of the model. In Ref.
\cite{KNS}, it was emphasized that an anomalous scaling dimension of the
gauge field, arising from matter-field fluctuations, changes the interaction
between monopoles from $1/r$ to $\ln r$
in three dimensions.  It was then argued that a
monopole-antimonopole unbinding transition similar to the Kosterlitz-Thouless
(KT) transition takes place, but now in three dimensions. From this, we
concluded that test charges undergo a deconfinement transition.

It must be pointed out that the authors
of Refs. \cite{FradShe,Nagaosa}, were looking for a transition similar to those
encountered in $d=3+1$,
namely ordinary first- or second-order  phase transitions
\cite{FradShe}. In Ref. \cite{Nagaosa},  a duality transformation was performed
showing that the disorder parameter $\langle\phi_V\rangle$ is always different
from zero, implying that $\langle\phi\rangle$ is always zero.
{\it This result is essentially correct and is perfectly consistent with the scenario in}  Ref. \cite{KNS}
and explained further in the present paper.

A main result in our letter \cite{KNS}
is that there exists
a non-trivial infrared stable fixed point in the theory in $d=2+1$
which drives the deconfinement transition.
There the
 anomalous
dimension of the gauge field is given
 by $\eta_A=1$ in
$d=2+1$ \cite{Herbut,Hove}.
{\it This result is exact as a consequence of gauge invariance. It implies that
the non-trivial infrared fixed point arises at an infinite bare gauge coupling.}
To see this, consider  the
boson-fermion Lagrangian ${\calL}={\calL}_f+{\calL}_b+{\calL}_M$.
Due to gauge invariance, the gauge coupling renormalizes to
$e^2=Z_A e_0^2$, where $Z_A$ is the
  wave function renormalization constant
of the  gauge field.
 The renormalization group (RG) $\beta$
function for the renormalized dimensionless gauge coupling
$\alpha=e^2/\mu$ has the following exact form in $2+1$ dimensions

\begin{equation}
\label{betaalpha}
\beta_\alpha(\alpha,g)=\mu\frac{\partial\alpha}{\partial\mu}
=[\gamma_A(\alpha,g)-1]\alpha,
\end{equation}
where $g$ is the renormalized dimensionless $|\phi|^4$
coupling and $\gamma_A=\mu\partial\ln Z_A/\partial\mu$.
Let us assume that there exist non-trivial infrared stable fixed
points $\alpha_*$ and $g_*$, where
the $ \beta $ functions
$\beta_\alpha$ and $\beta_g$ vanish. We have explained in
Ref. \cite{KNS}
why such fixed points must exist.
(For similar arguments, see Ref.~\cite{Kovner}). Moreover,
large-scale Monte-Carlo simulations have demonstrated explicitly
the existence of such a non-trivial fixed point \cite{Hove,Nguyen}
(see also Ref. \cite{Olsson}). Its existence
has long been assured theoretically by duality arguments
\cite{Dasgupta,Kleinert} (see also Section II-B).
We shall 
 not repeat the arguments
and details here. Instead, we focus on the physical consequences of
the non-trivial fixed point.

We would like to stress an important point, pertinent to $d=2+1$
dimensions, and quite
different from the situation for  $d=3+1$.
As $\alpha\to \alpha_*$,
the bare coupling
$e_0^2$ must tend to infinity.
 {\it By definition}, the above $\beta$ function
is given at fixed $\Lambda$, $\alpha_0$, and $g_0$. Here,
$\Lambda$ is the ultraviolet cutoff while $\alpha_0=e_0^2/\Lambda$
and $g_0=u_0/\Lambda$ are the dimensionless {\it bare} couplings.
The fixed point is reached for $\mu\to 0$. Alternatively,
the fixed point is reached for $\Lambda\to\infty$
if
$\mu$ is held fixed. However, since $\alpha_0$ is
fixed it follows that $e_0^2\to\infty$ as $\Lambda\to\infty$.
Thus, in $d=2+1$, the fixed point theory is at {\em infinite
bare gauge  coupling}. One might object
that this infinite gauge coupling
cannot be relevant
for the cuprates which have an infinite value of $e_0^2$
at {\it all} scales, not only in the scale invariant
regime. This is true, but
irrelevant as far as the
 deconfinement
transition is concerned, which is determined by the
non-trivial fixed point structure.
 The situation  is analogous to the $O(N)$ non-linear
$\sigma$ model as opposed to the $O(N)$ $\phi^4$ model.
These models are quite different, but agree with each other
at the critical point \cite{KS,ZJ}, thus
belonging to the same universality class.
In our case, the model with the Maxwell term at the fixed
point has the
same correlation functions as the model without it
also at the fixed point.

To summarize the discussion in the above paragraph, the non-compact
action with no Maxwell term has the same critical behavior as the
compact one {\it at the critical point corresponding to
a non-trivial fixed point, characterized by an infinite
bare coupling}. Had we started from an infinitely weak bare coupling,
the only fixed point we would have any hope of reaching for
 $d=2+1$
would be the Gaussian fixed point.

In Ref.  \cite{KNS}
we have pointed out that chiral symmetry
breaking can destroy the deconfinement in the fermionic case.  We
want to point out that for the combined boson-fermion model,
${\calL}={\calL}_f+{\calL}_b+{\calL}_M$,
chiral symmetry breaking does not spoil the deconfinement transition.
Chiral symmetry breaking occurs at a lower value of number of
fermion flavours $N_f$, when also bosons are present. Kim and Lee
\cite{Kim} claimed that the critical value of $N_f$ is decreased
by a factor two. Since we have typically $N_f^c\sim 3$ and the
physical number of fermion components in the cuprates is $N_f=2$,
Kim and Lee argued that spin-charge separation would occur at
finite doping \cite{Kim}.

\subsection{Anomalous scaling and the potential between
test charges}

The high-energy physics literature is usually concerned with 
 $d=4$
 and use low-dimensions
 only in toy models. In condensed matter
physics, however, $2+1$-dimensional gauge theories are supposed to
describe real physical phenomena such as the anomalous properties of
high-$T_c$ superconductors \cite{IL}, or the physics of QHA
\cite{Affleck,Sachdev}. For $d\in(2,4]$ the gauge
coupling $\beta$-function may be written as

\begin{equation}
\label{betaalphad}
\beta_\alpha(\alpha,g)=[\gamma_A(\alpha,g)+d-4]\alpha.
\end{equation}
Non-trivial fixed points induce an anomalous scaling behavior
in the gauge field propagator. In the Landau gauge we have that

\begin{equation}
D_{\mu\nu}(p)=D(p)\left(\delta_{\mu\nu}-\frac{p_\mu p_\nu}{p^2}\right),
\end{equation}
with the large distance behavior given by

\begin{equation}
\label{D}
D(p)\sim\frac{1}{|p|^{2-\eta_A}}.
\end{equation}
The anomalous scaling dimension is given exactly by \cite{Herbut,Hove}

\begin{equation}
\label{etaA}
\eta_A\equiv\gamma_A(\alpha_*,g_*)=4-d.
\end{equation}
Due to the above result, the propagator  (\ref{D}) in configuration  space becomes

\begin{equation}
\label{Dspace}
D(x)\sim\frac{1}{|x|^{d-2+\eta_A}}\sim\frac{1}{|x|^2},
\end{equation}
for all $d\in(2,4]$.
The potential between {\it effective electric charges $q(R)$},
separated by a large distance $R$ in $(d-1)$-dimensional space is given by

\begin{equation}
\label{potential}
V(R)\sim\frac{q^2(R)}{R^{d-3}},
\end{equation}
where

\begin{equation}
\label{effcharge}
q^2(R)\sim \frac{1-(\Lambda R)^{-\eta_A}}{\eta_A}\sim
\frac{(\Lambda R)^{d-4}-1}{d-4},
\end{equation}
and where $\Lambda$ is a short distance cutoff. The anomalous scaling in Eq.
(\ref{effcharge}) is a consequence of the coupling to matter fields.
Due to it, the potential $V(R)$ behaves effectively like $1/R$
for $d=3$. For $d=4$, it goes like $\ln (\Lambda R)/R$,
 while for $d=2$,
it has a confining behavior
proportional to
 $R$.   The regime
governed by the Gaussian  fixed point has
$q^2(R)=q_0^2={\rm const}$,
and
 corresponds
to the so called Coulomb phase. In this phase, the four-dimensional theory
has  $V(R)=q_0^2/R$, whereas
$V(R)=q_0^2\ln R$  for $d=3$.
{\it We see that the
non-trivial infrared behavior
induces an effective electric potential between
test charges similar to that which characterizes the Coulomb phase
in $d=4$}.
If we
extrapolate to $d=2$, we obtain  $V(R)=q_0^2 R$. Note that in $d=2$, we
obtain a confining potential irrespective of whether anomalous scaling
is taken into account or not.

In compact abelian gauge theories a confined phase is realized  by
the formation
of
{\it electric} flux tubes connecting electric charges.
These flux tubes
are the dual analogs
of the {\it magnetic} flux tubes connecting
magnetic monopoles \cite{tHooft,Conf}. There is a Dirac relation between the effective
electric and magnetic charges

\begin{equation}
q(R)q_m(R)\sim 1.
\end{equation}

Let us consider now the potential between the magnetic charges

\begin{equation}
\label{magpoten}
V_m(R)\sim\frac{q_m^2(R)}{R^{d-3}}\sim\frac{1}{q^2(R)R^{d-3}}.
\end{equation}
From Eq. (\ref{effcharge}) we see that for $d=4$ the
magnetic potential behaves like $1/[R\ln(\Lambda R)]$.
However, for $d=3$ we have

\begin{equation}
V_m(R)\sim \frac{1}{R},
\end{equation}
which is self-dual with respect to the potential between
electric test charges.

The Higgs phase for the {\it electric charges} corresponds to $V(R)\sim {\rm const}$
because of the gauge field mass gap. The Higgs phase for {\it magnetic test charges},
on the other hand, is given by $V(R)\sim R$.  In the electric-magnetic duality
picture \cite{tHooft,Conf}  this Higgs phase for magnetic charges is exchanged by
the confined phase for electric charges. This scenario should be valid for
matter fields in the {\it adjoint} representation.
In the absence of matter fields, a compact $2+1$ dimensional gauge theory is
definitely confined permanently \cite{Polyakov}. The above result shows that
if matter
fields are present, a deconfined phase is also possible.
However, if the
matter fields are in the fundamental representation, the situation is
controversial \cite{ES,Nagaosa,Nayak,Matsui,Nayak1,KNS}. 
Our recent results in Ref. \cite{KNS}  
seem to be confirmed
by the Monte Carlo work in Ref. \cite{Chernodub1}.
The main purpose of this paper is to give more details on the
scenario proposed in Ref. \cite{KNS} and
to describe a theory for a deconfinement transition
in Abelian gauge theories coupled to matter
fields in the fundamental representation.

\subsection{Outline of the paper}
In Section II, we consider the lattice duality transformations to the
$2+1$-dimensional Abelian Higgs lattice (AHL) model, first the non-compact
case and later the compact case. We then discuss the possible ordinary
first- or second-order phase transitions these models can have, with
matter fields in the fundamental representation for the compact case.

In Section III, we construct the continuum effective Lagrangian and its
dual counterpart for the compact $2+1$-dimensional AHL model when
matter-fields have been integrated out.
Because these are central results of the paper, it behooves
us to announce them here.

The dual field theory  is given by  Eq. (\ref{CPSG}). It represents a description
of a three dimensional gas of point charges interacting with a {\it logarithmic}
pair-potential, given by Eq. (\ref{monopaction}). We emphasize that
the $3d$ $\ln$-plasma action of Eq. (\ref{monopaction}) emerges
from an underlying matter-coupled gauge theory, Eq. (\ref{compdual2}),
by integrating out the fluctuating matter fields and considering the
influence of {\it critical} matter fluctuations on the gauge-field
propagator. The result of this procedure is the effective theory
Eq. (\ref{LA}). Such matter-field
fluctuations endow the gauge-field propagator with an anomalous scaling
dimension $\eta_A = 4-d$ \cite{Herbut,Hove} which
in three dimensions alters the interaction
between the monopole configurations of the gauge-field from a
Coulomb-interaction $1/R$ to a $\ln R$ interaction.

Recall that in contrast to this, in the classic
treatment by Polyakov \cite{Polyakov} of compact three-dimensional QED with no
matter fields,  the standard  three-dimensional sine-Gordon field theory with a
quadratic gradient term,  describing the three dimensional Coulomb gas, is obtained.
This action is given by, in the notation of
Eq. (\ref{CPSG})
\begin{equation}
\label{SG}
S_{\rm SG}=\frac{1}{2t}\int d^3 x[\varphi(-\partial^2)
\varphi-2z_0\cos \varphi].
\end{equation}
Polyakov has demonstrated \cite{Polyakov} that Eq. (\ref{SG}) has no phase
transition, i. e. it is always massive. Our Eq. (\ref{CPSG}) differs
drastically from Eq. (\ref{SG}), due the presence of an anomalous gradient
term.

In the first part of section IV, we show using the Callan-Symanzik equations,
that the effective dual Lagrangian Eq. (\ref{CPSG}) has a massless and a massive
phase separated at a critical coupling $t_c$. Hence  a phase transition must
exist. This does not by itself suffice to show precisely {\it what sort of phase
transition} the system undergoes, nor does it allow us to construct the
correct flow
diagram of the coupling constants of the problem. It does, however, suffice to
show that two different phases exist.
Since the propagator of the problem is logarithmic in $d=2+1$,
a Hohenberg-Mermin-Wagner theorem \cite{HMW} holds. Under such
circumstances, it is very natural to conjecture
that any phase transition in the system, if it exists, must be of a
{\it topological character}.
In the second part of Section IV, we
construct the renormalization group flow equations for the problem and show
that the phase transition is of a KT-like type.

In Section V, we consider the connection between the renormalization group functions
obtained directly from the Abelian Higgs model, and the KT phase transition we
find in Section IV. The main point here is that we can use the value of the critical
coupling of the dual effective Lagrangian for the topological defects of the gauge
field to fix an {\it a priori} arbitrary constant which enters into evaluating critical
exponents for the {\it non-compact} Abelian Higgs model.

In Section VI, we conclude with a summary and outlook. Appendix A discusses another
type of sine-Gordon theory also exhibiting  a KT-like transition in three dimensions.
In Appendix B, we derive the flow equations for the stiffness parameter
and the fugacity of the system defined by Eq. (\ref{monopaction}), and of which
Eq. (\ref{CPSG}) is a field theory formulation.
In Appendix C, we compute the screened effective potential between
charges in the insulating phase of the $3d$ $\ln$-plasma. In Appendix D, for
completeness, we derive the exact equation of state for a $d$-dimensional
$\ln$-plasma {\it with no short-distance cutoff} and relate the singularities
in this plasma to the Callan-Symanzik approach of Section IV.
In Appendix E, we consider, also for completeness, the duality transformation
of the AHL model with a Chern-Simons term added. This case is of interest in
the fractional quantum Hall  effect \cite{Fradkin_book} and chiral spin
liquids \cite{Wen}.

\section{Duality in the abelian Higgs lattice model}

In this section we review the duality approach to the
AHL model. Although this is a well studied topic
\cite{Peskin,Banks,ES,Dasgupta,Kleinert}, it is worth  reviewing
it here in order to emphazise the differences and similarities
between the non-compact and compact cases. In particular, we
shall discuss the extent to which these cases exhibit ordinary
first- or second-order phase transitions.
The interesting case
including a Chern-Simons term will be discussed in Appendix D.

The essential point is that starting from a non-compact
or compact AHL model, the  dual action has  the general form

\begin{equation}
\label{dualaction}
S_{\rm dual}=\frac{1}{2}\sum_{i,j}h_{i\mu}M_{\mu\nu}({\bf r}_i-
{\bf r}_j)h_{i\nu}-i 2\pi\sum_i {\bf l}_i\cdot{\bf h}_i,
\end{equation}
where $h_{i\mu}\in(-\infty,\infty)$ and
${\bf l}_i$ are integer dual link variables. In the
non-compact case ${\bf l}_i$ satisfy the constraint
\begin{equation}
\nablab\cdot{\bf l}_i=0,
\end{equation}
whereas in the compact case, the right-hand side is nonzero
\begin{equation}
\nablab\cdot{\bf l}_i=Q_i,
\end{equation}
due to monopole charges $Q_i\in\mathbb{Z}$.
The symbol $\nablab$ denotes the gradient vector ona simple
cubic lattice of unit spacing
 with
components $\nabla_\mu f_i\equiv f_{i+\hat{\mu}}-f_i$.

\subsection{The non-compact case and the ``inverted'' $XY$
transition}

In the non-compact case, the partition function of the AHL model is
given by
\begin{equation}
Z=\sum_{\{n_{i\mu}\}}
\int_{-\pi}^\pi\left[\prod_i\frac{d\theta_i}{2\pi}\right]
\int_{-\infty}^\infty\left[\prod_{i,\mu}dA_{i\mu}\right]
\exp(-S),
\label{ncaction}
\end{equation}
where the action $S$ is given by the Villain approximation
\begin{eqnarray}
\label{non-compAHL}
S&=&
\frac{\beta}{2}\sum_{i,\mu}(\nabla_\mu\theta_i-A_{i\mu}-2\pi n_{i\mu})^2
+\frac{1}{2e^2}\sum_i(\nablab\times{\bf A}_i)^2.  \nonumber \\&&
\end{eqnarray}
Using the identity
\begin{equation}
\label{identity}
\sum_{m=-\infty}^\infty e^{(-t/2)m^2+ixm}=
\sqrt{\frac{2\pi}{t}}\sum_{n=-\infty}^{\infty}e^{(-1/2t)(x-2\pi n)^2},
\end{equation}
following directly from
Poisson's
formula
\begin{equation}
\sum_{n=-\infty}^\infty F(n)=\sum_{m=-\infty}^\infty
\int_{-\infty}^\infty dx F(x) e^{2\pi imx},
\label{@Poi}\end{equation}
we obtain
\begin{eqnarray}
\label{Z1}
Z&=&\int_{-\infty}^\infty\left[\prod_{i,\mu}dA_{i\mu}\right]
\sum_{\{{\bf m}_i\}}\delta_{\nablabs\cdot{\bf m}_i,0}
\exp\left\{\sum_i\left[-\frac{1}{2\beta}{\bf m}_i^2\right.\right.
\nonumber\\
&+&\left.\left.i {\bf A}_i\cdot
{\bf m}_i-\frac{1}{2e^2}(\nablab\times{\bf A}_i)^2
\right]\right\}.
\end{eqnarray}
The Kronecker delta in Eq. (\ref{Z1}) is generated by the
$\theta_i$ integrations. Now we should integrate out the
gauge field ${\bf A}_i$. The easiest way of performing
this integration is by the introduction of an auxiliary
field ${\bf h}_i$ such that the partition function can
be rewritten as

\begin{eqnarray}
\label{Z2}
Z&=&\int_{-\infty}^\infty\int_{-\infty}^\infty\int_{-\infty}^\infty
\left[\prod_{i,\mu}dA_{i\mu}dh_{i\mu}db_{i\mu}\right]
\sum_{\{{\bf M}_i\}}\delta(\nablab\cdot{\bf b}_i)
\exp\left\{\sum_i\left[-\frac{1}{2\beta}{\bf b}_i^2
\right.\right.\nonumber\\
&+&\left.\left.i {\bf A}_i\cdot
({\bf b}_i-\nablab\times{\bf h}_i)-\frac{e^2}{2}{\bf h}_i^2
+2\pi i {\bf M}_i\cdot{\bf b}_i
\right]\right\},
\end{eqnarray}
where a summation by parts has been done to replace
${\bf h}_i\cdot(\nablab\times{\bf A}_i)$ by
${\bf A}_i\cdot(\nablab\times{\bf h}_i)$,
and we have used the
Poisson formula
(\ref{@Poi})
to replace the integer variables ${\bf m}_i$ by
continuum variables
${\bf b}_i$, at the cost of an additional sum over integer
variables ${\bf M}_i$.
We may now integrate out ${\bf A}_i$ to obtain a
delta function $\delta({\bf b}_i-\nablab\times{\bf h}_i)$, after which
also
${\bf b}_i$ can be
 integrated out ${\bf b}_i$, yielding
\begin{eqnarray}
\label{Z3}
Z&=&\sum_{\{{\bf M}_i\}}\int_{-\infty}^\infty\left[\prod_{i,\mu}
dh_{i\mu}\right]\exp\left\{-\sum_i\left[
\frac{1}{2\beta}(\nablab\times{\bf h}_i)^2
\right.\right.\nonumber\\
&+&\left.\left.
\frac{e^2}{2}{\bf h}_i^2-2\pi i {\bf M}_i\cdot(\nablab\times{\bf h}_i)
\right]\right\}.
\end{eqnarray}
Summing the last term
 in the exponent
by parts
 and    going over
to integer variables ${\bf l}_i=\nablab\times{\bf M}_i$,
we obtain
\begin{eqnarray}
\label{Z4}
Z&=&\sum_{\{{\bf l}_i\}}\int_{-\infty}^\infty\left[\prod_{i,\mu}
dh_{i\mu}\right]\delta_{\nablabs\cdot{\bf l}_i,0}\exp\left\{-\sum_i\left[
\frac{1}{2\beta}(\nablab\times{\bf h}_i)^2
\right.\right.\nonumber\\
&+&\left.\left.
\frac{e^2}{2}{\bf h}_i^2-2\pi i {\bf l}_i\cdot{\bf h}_i
\right]\right\}.
\end{eqnarray}
Note that the Kronecker delta constraint above is a direct
consequence of our change to integer-valued  variables. If
${\bf h}_i$ is integrated out we obtain

\begin{equation}
\label{dual}
Z=Z_0\sum_{{\bf l}_i}
\delta_{\nablabs\cdot{\bf l}_i,0}
\exp\left[-2\pi^2\beta\sum_{i,j,\mu}l_{i\mu}
D({\bf r}_i-{\bf r}_j)l_{j\mu}\right],
\end{equation}
where the Green function $G$ has the large-distance
behavior
\begin{equation}
D({\bf r}_i-{\bf r}_j)\sim
\frac{e^{-\sqrt{\beta} e|{\bf r}_i-{\bf r}_j|}}{4\pi|{\bf r}_i-{\bf r}_j|}.
\end{equation}
The factor $Z_0$ in Eq. (\ref{dual}) corresponds to the partition
function of a free massive gauge boson theory.

Eq. (\ref{dual}) is the dual representation of the partition
function for the non-compact AHL model. Due to the constraint
$\nablab\cdot{\bf l}_i=0$,
the integer links ${\bf l}_i$ form
closed loops.

By taking the limit $e\to 0$ in Eq. (\ref{Z1}), we obtain

\begin{equation}
\label{loopgas1}
Z|_{e=0}=\sum_{\{{\bf m}_i\}}\delta_{\nablabs\cdot{\bf m}_i,0}
\exp\left(-\frac{1}{2\beta}\sum_i{\bf m}_i^2\right),
\end{equation}
which is the loop gas representation of the $XY$ model.
If, on the other hand, we take the limit $\beta\to\infty$
in Eq. (\ref{dual}), we obtain the loop gas representation
of the ``frozen superconductor'' \cite{Peskin}

\begin{equation}
\label{loopgas2}
Z|_{\beta=\infty}=\sum_{\{{\bf l}_i\}}\delta_{\nablabs\cdot{\bf l}_i,0}
\exp\left(-\frac{2\pi^2}{e^2}\sum_i{\bf l}_i^2\right),
\end{equation}
which has precisely the same form as in Eq. (\ref{loopgas1}).
Therefore, the $XY$ model is equivalent to the frozen
superconductor, provided the Dirac like relation
$e^2=4\pi^2\beta$ holds.  Eq. (\ref{dual}) is a reformulation
of Eq. (\ref{ncaction}) in terms of the topological defects
of the model, which are identified as integer-valued vortex
strings forming closed loops.

If we consider the phase diagram in the $e^2-T$-plane (with
$T=1/\beta$), we can use
Eqs. (\ref{loopgas1}) and (\ref{loopgas2}) to establish the critical
points on the axes $e^2$ and $T$, corresponding to $T\to 0$ and
$e^2\to 0$ limits, respectively. From Eq. (\ref{loopgas1}) we see
that when $e^2\to 0$ we have a $XY$ critical point on the $T$-axis.
Eq. (\ref{loopgas2}) has exactly the same form as Eq. (\ref{loopgas1}),
but corresponds to the $T\to 0$ limit. The critical point
in this limit is therefore $e_c^2=4\pi^2/T_c$, with $T_c$ being the
critical temperature of the $XY$ transition as described by the
Villain approximation. This is the so called ``inverted'' $XY$
transition ($IXY$) \cite{Dasgupta}. From the existence of these
two critical points we can establish a phase diagram where
there is a critical line connecting them \cite{Dasgupta}.
The ordered superconducting phase corresponds to the region
$0<e^2<e_c^2$.

\subsection{The compact case and the absence of an ordinary
phase transition}

In the compact AHL model the gauge field $A_{i\mu}\in[-\pi,\pi]$. The
corresponding Villain action is now given by

\begin{equation}
\label{compactAH}
\tilde{S}=\frac{\beta}{2}\sum_i(\nabla_{\mu}\theta_i-
A_{i\mu}-2\pi n_{i\mu})^2
+\frac{1}{2e^2}\sum_i(\epsilon_{\mu\nu\lambda}\nabla_\nu A_{i\lambda}
-2\pi N_{i\mu})^2,
\end{equation}
and in the partition function we should sum over both
integers $n_{i\mu}$ and $N_{i\mu}$. Using the identity
(\ref{identity}) we obtain

\begin{equation}
\label{compZ1}
Z=\sum_{\{{\bf n}_i\}}\sum_{\{{\bf m}_i\}}
\int_{-\pi}^\pi\left[\prod_{i,\mu}
\frac{dA_{i\mu}}{2\pi}\right]\int_{-\pi}^\pi\left[\prod_{i}
\frac{d\theta_{i}}{2\pi}\right]\exp(S'),
\end{equation}
where

\begin{equation}
\label{Sprime}
S'=\sum_i\left[\frac{1}{2\beta}{\bf n}_i^2+i {\bf n}_i\cdot
(\nabla\theta_i-{\bf A}_i)
+\frac{e^2}{2}{\bf m}_i^2+i{\bf m}_i\cdot(\nablab\times{\bf A}_i)
\right].
\end{equation}
Now we integrate out $A_{i\mu}$ and $\theta_i$ to obtain

\begin{eqnarray}
\label{compZ2}
Z&=&\sum_{\{{\bf n}_i\},\{{\bf m}_i\}}
\delta_{\nablabs\cdot{\bf n}_i,0}\delta_{\nablab\times{\bf m}_i,{\bf n}_i}
\exp\left[-\sum_i\left(\frac{1}{2\beta}{\bf n}_i^2
+\frac{e^2}{2}{\bf m}_i^2\right)\right]
\nonumber\\
&=&\sum_{\{{\bf m}_i\}}\exp\left[-\sum_i\left(\frac{1}{2\beta}(
\nablab\times{\bf m}_i)^2
+\frac{e^2}{2}{\bf m}_i^2\right)\right]\nonumber\\
&=&\sum_{\{{\bf l}_i\}}\int_{-\infty}^\infty\left[\prod_{i,\mu}dh_{i\mu}
\right]\exp\left\{-\sum_i\left[\frac{1}{2\beta}(\nablab\times
{\bf h}_i)^2\right.\right.
+\left.\left.\frac{e^2}{2}{\bf h}_i^2-2\pi i{\bf l}_i\cdot{\bf h}_i
\right]\right\},
\end{eqnarray}
where from the second to the third line we used the Poisson
formula. Note the difference between Eq. (\ref{compZ2}) and
its non-compact counterpart Eq. (\ref{Z4}). In the latter there is a
Kronecker delta constraint $\nablab\cdot{\bf l}_i=0$ while in the
former there is no such a constraint. As we shall see, this difference
has important consequences. We proceed by integrating out $h_{i\mu}$,
thus obtaining the partition function

\begin{equation}
\label{compdual1}
Z=Z_0\sum_{\{{\bf l}_i\}}\exp\left[-2\pi^2\beta\sum_{i,j}
l_{i\mu}D_{\mu\nu}({\bf r}_i-{\bf r}_j)l_{j\nu}\right],
\end{equation}
where

\begin{equation}
\label{Gmunu}
D_{\mu\nu}({\bf r}_i-{\bf r}_j)
=\left(\delta_{\mu\nu}-\frac{\nabla_\mu\nabla_\nu}
{\beta e^2}\right)D({\bf r}_i-{\bf r}_j),
\end{equation}

\begin{equation}
(-\nabla^2+\beta e^2)D({\bf r}_i-{\bf r}_j)=\delta_{ij}.
\end{equation}
Due to the constraint $\nablab\cdot{\bf l}_i=0$, the term containing
$\nabla_\mu\nabla_\nu$ in Eq. (\ref{Gmunu}) does not contribute in
the non-compact case, and Eq. (\ref{dual}) results. In the compact
case, on the other hand,  $\nablab\cdot{\bf l}_i$ is completely
unconstrained and can take any integer value. Thus,
in order to bring out the differences and similarities between
Eqs. (\ref{compdual1}) and (\ref{dual}), and also to
identify the character of the topological defects of Eq. (\ref{compactAH})
appearing in Eq. (\ref{compdual1}), we can
introduce an auxiliary integer-valued scalar field $Q_i$ such that
$\nablab\cdot{\bf l}_i=Q_i$ and rewrite the partition function
(\ref{compdual1}) as

\begin{equation}
\label{compdual2}\!\!\!\!
~~~~~~~
Z=Z_0\sum_{\{{\bf l}_i\}}\sum_{\{Q_i\}}\delta_{\nablabs\cdot{\bf l}_i,Q_i}
\exp\left[ -2\pi^2\beta\,\sum_{i,j}
D({\bf r}_i\!-\!{\bf r}_j)\left(l_{i\mu}l_{j\mu}
\!+\!\frac{1}{{e^2\beta}}Q_i Q_j\right)\right] . \!\!\!\!
\end{equation}
Whereas the non-compact theory has only  closed vortex lines as topological
defects, the compact case contains also open lines with integer-valued monopoles
of charge $Q_i$ at the ends.

In the limit $\beta\to 0$, Eq. (\ref{compdual2}), the vortex loops are frozen out
and (\ref{compdual2}) is the dual representation of three-dimensional lattice
compact QED \cite{Polyakov} describing
a Coulomb gas of monopoles in three dimensions. This is equivalent
to a sine-Gordon model which is always massive in three
dimensions, and leads to the well known result that compact
QED in three dimensions has permanent confinement of electric
charges, since the monopole gas will always be in the plasma
phase. As shown by Polyakov \cite{Polyakov}, we obtain as a
consequence that the Wilson loop satisfies the area law.

As in the non-compact case, the limit $e^2\to 0$ corresponds
to the Villain form of the $XY$ model. Thus, if we consider
again a phase diagram in the $(e^2,T)$-plane we have that
a critical point at $T_c$ exists on the $T$-axis. However,
as we shall now show, {\it there is no $IXY$ transition in the
compact case}. To see this, let us take the ``frozen'' limit
$\beta\to\infty$ in Eq. (\ref{compdual2}). The result is

\begin{equation}
\label{frozcomp}
Z=Z_0\sum_{\{{\bf l}_i\}}\sum_{\{Q_i\}}\delta_{\nablabs\cdot{\bf l}_i,Q_i}
\exp{\left(-\frac{2\pi^2}{e^2}\sum_i{\bf l}_i^2\right)}.
\end{equation}
The sum over $Q_i$ is trivially done,
$\sum_{\{Q_i\}}\delta_{\nablabs\cdot{\bf l}_i,Q_i}=1$ after which there is
no constraint. We are left with a trivial sum over ${\bf l}_i$
giving Jacobi $\vartheta$-functions $\vartheta_3(0,e^{-2\pi^2/e^2})$.
Since this function is analytic, there is no phase transition on the
$e^2$-axis, in contrast to the non-compact case. Thus, at first
sight it seems that there is no phase transition in the compact
AHL model with matter fields in the fundamental representation,
except for the $XY$-transition on the $T$-axis. That is, there
appears to be no ordinary second- or first-order phase transition
in the interior of the phase diagram of this model.
However, in the next Sections we shall derive an effective
Lagrangian for the compact Abelian Higgs model in $2+1$ dimensions,
which will
be shown to nevertheless exhibit a topological phase transition
of the KT type.

\section{Effective Lagrangian}

This section is one of the central parts of the paper in which  we
shall derive an effective field theory for the compact
Abelian Higgs model in $d=2+1$
dimensions. More precisely, we derive a continuum action, Eq. (\ref{CPSG})
below,  for the dual model of the system, obtained after matter fields
have been integrated out leaving an effective theory for the monopoles
of the problem. It will turn out that the effective dual lagrangian for
the $2+1$-dimensional compact Abelian Higgs
model, is described by a theory which has
many similarities to the sine-Gordon theory of Polyakov's pure compact
electrodynamics in $d=2+1$ \cite{Polyakov}. The crucial difference lies
in the fact that the gradient term in the dual theory receives an
anomalous dimension after the matter-fields have been integrated out.
It is the presence of this anomalous gradient term induced by matter-field
fluctuations which eventually will lead to the possibility of a
deconfinement transition in $d=2+1$, in contrast to the classical
Polyakov-result of permanent confinement pertaining to the pure gauge theory .

\subsection{Three-dimensional compact QED}

Let us consider the Euclidean Maxwell action in three dimensions:

\begin{equation}
\label{Maction}
S=\int d^3x \frac{1}{4e^2}F_{\mu\nu}^2,
\end{equation}
where
$F_{\mu\nu}=\partial_\mu a_\nu-\partial_\nu a_\mu$.
In order to account for monopoles, we
have to subtract
from $F_{\mu \nu }$
the gauge field of monopoles \cite{Nato}
\begin{equation}
F^M_{\mu \nu }(x)=2\pi   \epsilon _{\mu \nu  \lambda } \delta _ \lambda (x;L),
\label{@}\end{equation}
where
$ \delta _ \lambda (x;L) $ is a delta function on lines $L$.
The dual field strength of $ \tilde F_ \lambda^M
=
\epsilon _{\mu \nu  \lambda }F_{\mu \nu }^M/2
 $ has  divergences
at the end points of the lines $L$,
say \cite{Conf,Nato}
\begin{equation}
\partial _\mu\tilde F_\mu^M
\label{monopole}
=2\pi n(x)=2\pi\sum_i Q_i \,\delta^3(x-x_i),
\end{equation}
where $Q_i$ may be arbitrary are integers
counting the number of  lines ending at $x_i$.
The shape of the lines is physically irrelevant.
They are the Dirac strings of
the monopoles at $x_i$.
Under shape deformations,  $ F_{\mu \nu }^M$ undergoes the
monopole gauge transformations
$ F_{\mu \nu }^M\rightarrow
 F_{\mu \nu }^M+
\partial _\mu \Lambda _ \nu ^M
-\partial _\nu \Lambda _ \mu^M$
which leave $\tilde F_\mu^M$ invariant.

An ordinary  gauge transformation can be used to bring
$ F^M_{\mu \nu }(x)$ to the form
\begin{equation}
\label{F}
F^M_{\mu\nu}=-
2\pi\epsilon_{\mu\nu\lambda}\partial_\lambda
\int d^3y \frac{1}{4\pi|x-y|}n(y),
\end{equation}
whose dual field strength is

\begin{equation}
\label{dualF}
\tilde{F}_\mu^M=
-2\pi\partial_\mu\int d^3y \frac{1}{4\pi|x-y|} n(y).
\end{equation}
By substituting
$F_{\mu\nu}$
by
$F_{\mu\nu}-
F_{\mu\nu}^M$
in
Eq. (\ref{Maction}), we obtain the action

 \begin{equation}
\label{newMaction}
S=\int d^3x\frac{1}{4e^2}F_{\mu\nu}^2
+\frac{2\pi^2}{e^2}\int d^3x\int d^3y~ n(x)\frac{1}{4\pi|x-y|}n(y).
\end{equation}
The action
(\ref{newMaction}) corresponds to the continuum counterpart
of the $\beta\to 0$ limit of the lattice action in Eq.
(\ref{compdual2})
describing a Coulomb gas of monopoles. This is known to be
equivalent to a sine-Gordon action
as the one in Eq. (\ref{SG}). In three dimensions this theory is
always massive and it was shown by Polyakov \cite{Polyakov} that
this implies an area law for the Wilson loop. Thus, electric test
charges in three-dimensional compact QED are permanently confined.

\subsection{Anomalous three-dimensional compact QED}

When bosonic matter fields are present, the topological defects
of the theory are vortex loops and vortex lines having monopoles
with opposite charges at the ends. The vortex lines connecting
the monopoles have a line tension $\sigma$ which vanishes as
the scalar bosons become massless. Thus, when the vortex lines
become tensionless, we are left with a gas of monopoles. However,
the anomalous scaling of the gauge field due to matter fields
alters the interaction between pair of monopoles with respect
to the ordinary Coulomb interaction case. This will lead us to the
{\it anomalous} Coulomb gas to be described below.

From the exact behavior of the critical gauge field propagator we have
discussed in Section IB, we can write an effective quadratic
{\it non-local} Lagrangian for the gauge field:

\begin{eqnarray}
\label{LA}
~~~~~~~~~~~~~~~~~{\calL}_A&=&\frac{K}{4}F_{\mu\nu}\frac{1}
{(-\partial^2)^{\eta_A/2}}F_{\mu\nu}
\nonumber\\
~~~~~~~~~~~~~~~~~&=&\frac{K}{2}\tilde{F}_{\mu}
\frac{1}{(-\partial^2)^{\eta_A/2}}\tilde{F}_{\mu},
\end{eqnarray}
where the constant $K=K(\alpha_*,g_*)$ and in the second line of
Eq. (\ref{LA}) we have rewritten ${\calL}_A$ in terms of the
dual field strength. Specializing to three dimensions, we have
$\tilde{F}_\mu=\epsilon_{\mu\nu\lambda}F_{\nu\lambda}/2$
and $\eta_A=1$. After introducing an auxiliary vector field
$b_\mu$, we obtain the equivalent Lagrangian:

\begin{equation}
{\calL}_A'=\frac{1}{2K}b_\mu\sqrt{-\partial^2}~b_\mu
+ib_\mu \tilde{F}_\mu.
\end{equation}
In order to take into account the monopoles, we use the expression
for $\tilde{F}_\mu$ as given in Eq. (\ref{dualF}). By introducing a
new field through $b_\mu=\partial_\mu\varphi$ and using integration by
parts, we obtain:

\begin{equation}
{\calL}_A''=\frac{1}{2K}(\partial_\mu\varphi)\sqrt{-\partial^2}
(\partial_\mu\varphi)+i 2\pi n(x)\varphi(x).
\end{equation}
Integrating out $\varphi$ and using Eq. (\ref{monopole}), we obtain
the monopole action:

\begin{equation}
\label{monopaction}
S_{\rm mon}=2\pi^2 K\sum_{i,j} Q_i Q_j G(x_i-x_j),
\end{equation}
where

\begin{equation}
G(x)=\int\frac{d^3 k}{(2\pi)^3}\frac{e^{ik\cdot x}}{|k|^3}.
\end{equation}
Thus, instead of having  a standard three-dimensional Coulomb gas with
interaction potentials 
$1/|x_i-x_j|$,
we have a three-dimensional gas  of point
particles of charge $ Q_i = \pm|Q|$ (with overall charge-neutrality,
see Section IVA) {\it with logarithmic interactions}, much akin to
the situation one has in two dimensions. We emphasize, once more, that
this is a result of integrating out matter-field fluctuations and
considering the effect of {\it critical} such fluctuations on the
gauge-field propagator, which is seen to acquire an anomalous
scaling dimension from these fluctuations, cf. Eq. (\ref{LA}).
 It therefore seems {\it plausible}, at the very least,
that one should consider the possibility of having a KT-transition
of unbinding of monopole-antimonopole pairs, but now
in three dimensions. If this turns out to be the case, then the
confinement-deconfinement transition in the $2+1$-dimensional
compact Abelian Higgs model with matter fields in the fundamental
representation, would be of a {\it topological} nature with no local
order parameter, consistent with previous work \cite{FradShe,Nagaosa}.

We are now ready to state one of the main results of this paper.
The system defined by Eq. (\ref{LA}) and (\ref{monopaction})
can be brought into the form of a
sine-Gordon theory, as in the two-dimensional case,
{\it but now with  an anomalous propagator}, whose action is

\begin{equation}
\label{CPSG}
S_{\rm ASG}=\frac{1}{2t}\int d^3 x[\varphi(-\partial^2)^{3/2}
\varphi-2z_0\cos \varphi],
\end{equation}
where $t=4\pi^2 K$ and $z_0=4\pi^2 K\zeta$, with $\zeta$ being the
fugacity of the Coulomb gas of monopoles. In Eq. (\ref{CPSG}),
$S_{\rm ASG}$ refers to the action of what we name the anomalous
sine-Gordon (ASG) theory, since the cubic power of the propagator arises
from the anomalous scaling dimension of the gauge field.
The manner in which the coupling constant  $t$ enters
in Eq. (\ref{CPSG}) shows that it regulates the stiffness of
the phase field $\varphi$. Since $t \propto K$, we shall in following
sections refer to $K$ as a stiffness parameter.

\section{Renormalization Group analysis of the
anomalous sine-Gordon model}

This section is another central part of the paper. Here, we shall
consider an exact scaling argument applied to Eq. (\ref{CPSG}).
The scaling argument will suffice to demonstrate that this model
has a phase transition. We emphasize this as an important point,
since recent numerical studies
\cite{Chernodub1} have provided strong support for the picture
proposed in Ref. \cite{KNS} that matter-field coupled to compact
$U(1)$ gauge fields in $d=2+1$ lead to a recombination of magnetic
monopoles into dipoles. For the dual electric charges, this
leads to a destruction of permanent confinement, and in Ref. \cite{KNS}
it was argued that this happened, not through any ordinary first or
second order phase transition, but rather through a
KT-like transition. The authors of Ref. \cite{Chernodub1} were
looking for more conventional phase transitions, concluding that none
were found, consistent with the results of Ref. \cite{KNS}.
Having established the existence of a phase transition, we
then go on to argue that it indeed  is of a KT-like type. The details
are as follows.

\subsection{Callan-Symanzik renormalization group  analysis}

Let us consider the renormalization of the anomalous sine-Gordon action
defined by Eq. (\ref{CPSG}). The infrared divergence is easily studied by
considering the cubic propagator $G(p)=1/|p|^3$ in real space. To this end,
we introduce an infrared cutoff $\mu$ as follows

\begin{eqnarray}
~~~G_\mu(x)&=&\int_{|p|>\mu}\frac{d^3p}{(2\pi)^3}\frac{e^{ip\cdot
 x}}{|p|^3}
=\frac{1}{2\pi^2}\left[\frac{\sin(\mu|x|)}{\mu|x|}
-{\rm ci}(\mu|x|)\right],
\end{eqnarray}
where ${\rm ci}(\lambda)$ is the cosine integral:

\begin{equation}
{\rm ci}(\lambda)\equiv -\int_\lambda^\infty \frac{\cos v}{v} dv.
\end{equation}
As $\mu\to 0$ we have

\begin{equation}
\label{expanG}
G_\mu(x)=\frac{1}{2\pi^2}[1-\gamma-\ln(\mu|x|)]+{\calO}(\mu),
\end{equation}
where $\gamma$ is the Euler-Mascheroni constant. For $x=0$, on the
other hand, $G_\mu(x)$ is ultraviolet divergent and becomes

\begin{equation}
\label{expanG1}
G_\mu(0)=\frac{1}{2\pi^2}\ln\left(\frac{\Lambda}{\mu}\right)
+{\rm const}+{\calO}(\frac{1}{\Lambda}),
\end{equation}
where $\Lambda$ is an ultraviolet cutoff.

Let us consider now the correlation function

\begin{eqnarray}
\left\langle\prod_{j=1}^n e^{iq_j \varphi(x_j)}\right\rangle
&=&\frac{1}{Z_0}\int{\calD}\varphi\exp\left\{-\frac{1}{2t}\int d^3 x
\left[\int d^3y~~\varphi(x) G_\mu^{-1}(x-y)\varphi(y)
\right.\right.\nonumber\\
&-&\left.\left.J(x)\varphi(x)\right]\right\},
\end{eqnarray}
where $J(x)=i\sum_j q_j\delta(x-x_j)$ and $Z_0$ is the above
functional integral for $J=0$.
Integrating out $\varphi$,
we obtain

\begin{equation}
\left\langle\prod_{j=1}^n e^{iq_j \varphi(x_j)}\right\rangle
=\exp\left[-\frac{1}{2}\sum_{i,j}G_\mu(x_i-x_j)q_i q_j\right].
\label{@nonzero}\end{equation}
Using Eqs. (\ref{expanG}) and (\ref{expanG1}), we obtain

\begin{eqnarray}
\label{Expansion}
\sum_{i,j}G_\mu(x_i-x_j)q_i q_j&=&-\frac{1}{2\pi^2}\left[
\left(\sum_i q_i\right)^2(\ln\mu+\gamma-1)
\right.\nonumber\\
&-&\left.\sum_iq_i^2\ln\Lambda+\sum_{i\neq j}
q_i q_j\ln|x_i-x_j|\right]+{\calO}(\mu).
\end{eqnarray}
Thus, as $\mu\to 0$ the only nonzero contributions to
(\ref{@nonzero}) satisfy the
 neutrality condition for the charge $\sum_i q_i=0$.
The expansion in Eq. (\ref{Expansion}) is essentially the same
as in the $d=2$ case, except for the $1/2\pi^2$ factor instead of
a $1/2\pi$, and minor differences in the constants.

The ultraviolet divergence of the phase field
$u_i(x)\equiv e^{iq_i\varphi(x)}$ is removed by introducing a wave
function renormalization $\zeta_i$ such that

\begin{equation}
u_i(x)=\zeta_i^{1/2}u_{i,R}(x),
\end{equation}
with $u_{i,R}$ being the renormalized counterpart of $u_i$ and

\begin{equation}
\zeta_i=\left(\frac{\Lambda}{\mu}\right)^{-q_i^2/(2\pi^2)}.
\end{equation}
Therefore, if we specialize to the case where $q_i=\pm|q|$ for all $i$,
the renormalized two-point correlation function is given by

\begin{equation}
\label{2pcorrelu}
\langle u_{i,R}(x)u_{i,R}^\dagger(0)\rangle\propto x^{-q^2/(2\pi^2)}.
\end{equation}
It follows that the dimension of $u_i$ is just $q^2/(4 \pi^2)$.

Due to the above analysis it is now easy to see how $z_0$
renormalizes in the ASG model. Note that the model is
super-renormalizable, just as the ordinary sine-Gordon model. Thus,
the renormalization of $z_0$ is achieved by taking
into account only tadpole contractions of $\cos\varphi$.
We obtain

\begin{equation}
\label{zrenorm}
z_0=Z_\varphi^{-1/2}z,
\end{equation}
where

\begin{equation}
Z_\varphi=\left(\frac{\Lambda}{\mu}\right)^{-t/(2\pi^2)}.
\end{equation}
Furthermore, we have the RG function

\begin{equation}
\label{etavarphi}
\eta_\varphi\equiv\mu\frac{\partial\ln Z_\varphi}{\partial\mu}=
\frac{t}{2\pi^2}.
\end{equation}
The renormalized $n$-point correlation function ${\calG}^{(n)}$ satisfies
the following Callan-Symanzik equation

\begin{equation}
\label{Callan-Symanzik}
\left(\mu\frac{\partial}{\partial\mu}+\frac{n}{2}\eta_\varphi
+\frac{1}{2}\eta_\varphi z\frac{\partial}{\partial z}
\right){\calG}^{(n)}(p_i,t,z)=0.
\end{equation}
Dimensional analysis, on the other hand,  gives

\begin{equation}
\label{DimenAnal}
\left[\mu\frac{\partial}{\partial\mu}+3z\frac{\partial}{\partial z}
+p_i\frac{\partial}{\partial p_i}+3(n-1)\right]{\calG}^{(n)}(p_i,t,z)=0,
\end{equation}
where $3(1-n)$ represents the
mass dimension of ${\calG}^{(n)}$. Using
Eq. (\ref{DimenAnal}) in (\ref{Callan-Symanzik}), we obtain

\begin{eqnarray}
\label{eqGn}
\left[p_i\frac{\partial}{\partial p_i}+3(n-1)-\frac{n}{2}\eta_\varphi
+\left(3-\frac{1}{2}\eta_\varphi\right)z\frac{\partial}{\partial z}
\right]{\calG}^{(n)}(p_i,t,z)=0.
\end{eqnarray}
For $p=0$ we have

\begin{equation}
(6-\eta_\varphi)z\frac{\partial{\calG}^{(n)}(0,t,z)}{\partial z}
=[6(1-n)+n\eta_\varphi]{\calG}^{(n)}(0,t,z),
\end{equation}
which gives the following scaling relation for small $z$

\begin{equation}
\label{scaling1}
{\calG}^{(n)}(0,t,z)\sim z^{[6(1-n)+n\eta_\varphi]/(6-\eta_\varphi)}.
\end{equation}
Also, it is clear from Eq. (\ref{eqGn}) that the scaling behavior of
the mass scale is

\begin{equation}
\label{scaling2}
m_\varphi\sim z^{2/(6-\eta_\varphi)}.
\end{equation}

The momentum space behavior of ${\calG}^{(2)}$ is
$\sim 1/p^{3-\eta_\varphi}$ and therefore ${\calG}^{(2)}$ becomes
singular in the ultraviolet if $3-\eta_\varphi<0$. This happens
for $t=6\pi^2$. For $t=6\pi^2$ the mass scale behaves like
$z^{2/3}$. This is an important difference between the
usual sine-Gordon model in two dimensions and the CPSG model in
three dimensions. The mass scale in the usual two-dimensional
sine-Gordon theory behaves linearly in $z$ when the singular
short-distance behavior is reached. There, this behavior is important
for the fermionization of the model, which establishes the equivalence
between the sine-Gordon model and the Thirring model in two dimensions
\cite{Coleman}.

From Eqs. (\ref{scaling1}) and (\ref{scaling2}) we see that
${\calG}^{(n)}(0,t,z)$ and $m_\varphi$ vanish for $t=t_c=12\pi^2$.
The interpretation of this result closely parallels the one
in the usual sine-Gordon model. For instance, it tells us that
at $t=t_c$ the operator $\cos\varphi$ is marginal, and
means that further renormalizations are necessary at
$t=t_c$. The situation exactly parallels the two-dimensional
case where a thorough analysis was carried out by Amit
{\it et al.} \cite{AGG}. For $t>t_c$ the anomalous sine-Gordon
model Eq. (\ref{CPSG}) is no longer renormalizable. These
results follow from the observation that the dimension of the
operator $\cos\varphi$ is just $\eta_\varphi/2$. Thus,
$\int d^3x\cos\varphi$ has dimension $\eta_\varphi/2-3$, which
means that $z$ has an effective dimension of
$3-\eta_\varphi/2$. Therefore, the interaction is relevant for
$\eta_\varphi<6$ or $t<t_c$, thus generating a mass. It is
marginal if $\eta_\varphi=6$ and irrelevant for $\eta_\varphi>6$,
or $t>t_c$, meaning that the theory is massless. Hence, there is
a phase where the field has a mass and another one where it is massless,
implying the existence of a genuine phase transition in the model
Eq. (\ref{CPSG}). This follows
from the fact that a mass changing from a finite value to zero on a
finite interval of coupling constants must do so in a non-analytic
fashion.  This conclusion is one of the main results of this paper.
Note, however, that since the above discussion is basically a spin wave
analysis and suffices to show that a phase transition exists, it does
not elucidate the {\it character} of the phase transition. In order to
understand the phase transition we have to account for the topological
defects in the theory \cite{KT}, and this is the purpose of the next
subsection.

\subsection{Kosterlitz-Thouless-like recursion relations for
the anomalous sine-Gordon model}

The above discussion strongly suggests the existence of a
phase-transition in the model defined by Eq. (\ref{CPSG}).
However, as we have already mentioned, the cosine interaction
becomes marginal at $t=t_c$. This means that it is not true
that $\beta(t)=0$ for all values of $t$. The analysis of the
previous subsection is neglecting the monopole fluctuations
which would lead to a renormalization of $t$. This situation
is well known for the logarithmic interaction in two dimensions and
leads to the KT recursion relations \cite{KT,Jose}.
Similar arguments can be used in our case.

Let us define the dimensionless coupling $y=z/\mu^3$. Using Eq.
(\ref{zrenorm}) we obtain the flow equation

\begin{equation}
\label{flowy1}
\mu\frac{\partial y}{\partial\mu}=\left(\frac{t}{4\pi^2}-3
\right)y.
\end{equation}
The above equation can be derived in another way, which is useful
for the purposes of this subsection. Let us consider again the
monopole action in Eq. (\ref{monopaction}). We can write
the partition function of the monopoles as

\begin{equation}
\label{partmon}
Z_{\rm mon}=\sum_{\{n(x)\}}\exp\left[-2\pi^2 K\int_{|x_i|>a}
d^3x\int_{|x_i'|>a}
d^3 x'
n(x)G(x-x')n(x')\right],
\end{equation}
where $a$ is a short distance cutoff.
Using Eqs. (\ref{expanG}) and (\ref{expanG1}), we rewrite the
above in the following form

\begin{eqnarray}
\label{partmon1}
Z_{\rm mon}&=&\sum_{\{n(x)\}}\mbox{'}\exp\left[-2\pi^2 K
\int_{|x_i|>a}d^3x\int_{|x_i'|>a} d^3 x'
n(x)\tilde{G}(x-x')n(x')
\right.\nonumber\\
&+&\left.\ln y_0\int_{|x_i|>a} d^3x n^2(x)\right],
\end{eqnarray}
where

\begin{equation}
\tilde{G}(x)=-\frac{1}{2\pi^2}\ln\frac{|x|}{a}.
\end{equation}
The prime on the
summation sign in Eq. (\ref{partmon1}) indicates that the
charge neutrality constraint implied by the large distance
limit is enforced. If we assume small $y_0$ such that configurations with
zero or one pair of monopoles are dominant, we obtain

\begin{equation}
\label{partmonapprox}
Z_{\rm mon}\approx 1 +y_0^2\int_{|x_i|>a}d^3x\int_{|x_i'|>a} d^3 x'
\frac{1}{|x-x'|^{2K}}.
\end{equation}
If we change the short distance cutoff in the integrals as $a\to ab$,
we see that the form
of Eq. (\ref{partmonapprox}) is
unchanged provided $x$ and $x'$ are rescaled
in such a way as to restore the previous integration
region and $y_0$ is changed according

\begin{equation}
y=y_0 b^{3-K}.
\end{equation}
If we define $l\equiv\ln b$, we obtain

\begin{equation}
\label{flowy2}
\frac{dy}{dl}=(3-K)y.
\end{equation}
Recalling that $t=4\pi^2K$, we see that
Eq. (\ref{flowy2}) is precisely Eq. (\ref{flowy1}), except for the sign,
which is due to differences in the cutoff procedure. Eq. (\ref{flowy2})
is analogous to the corresponding flow equation for the
fugacity in the ordinary KT transition \cite{KT,Jose}. In that
case we find instead $dy/dl=(2-\pi K)y$. The factor $2$ in the
usual KT case reflects the dimensionality. In our case we
have a factor $3$ instead (and also just $K$ rather than
$\pi K$).

It is also possible to derive recursion relations involving
the fugacity of the problem in arbitrary dimensions to
lowest orders in the fugacity for the $d$-dimensional Coulomb gas with
a power law interaction
\begin{equation}
\label{potential1}
V(x)=\frac{\Gamma\left(\frac{d-2}{2}\right)}{(4\pi)^{d/2}}
\left[\left(\frac{|x|}{a}\right)^{2-d}-1\right].
\end{equation}
This problem was considered by Kosterlitz
\cite{Kosterlitz}, who also
obtained the flow of the stiffness. The result is

\begin{equation}
\label{KTrec1}
\frac{dK^{-1}}{dl} = y^2 - (2-d) ~ K^{-1},
\end{equation}

\begin{equation}
\label{KTrec2}
\frac{d y}{dl} =  \left[d  - 2\pi^2f(d)K\right] ~ y,
\end{equation}
where $f(d)=(d-2)\Gamma[(d-2)/2]/(4\pi)^{d/2}$.
For $d=2$ this reduces to the KT flow equations.

However, we see that for $d=3$ the recursion
relations (\ref{KTrec1}) and (\ref{KTrec2}) do not have a fixed point and
therefore no phase transition happens in this case.
In  case of Eq. (\ref{monopaction}), on the other hand,
we have an anomalous Coulomb gas whose potential is
logarithmic in three dimensions. It is thus plausible
to conjecture that we would have a flow equation for
the stiffness similar to the $d=2$ KT case. As we shall see,
this is indeed the case.

For $d=3$, Eq. (\ref{KTrec2}) coincides with our Eq. (\ref{flowy2}).
However, since the potentials are different we should in fact
not expect the
same recursion relation for the fugacity. This suggests
that the ``spin wave'' picture of Section IV-A is not giving
the correct flow for the fugacity. Note that for
$d=2$
discussed in
 \cite{AGG}, {\it the spin wave analysis does in fact give the correct flow for the
fugacity}(see Appendix B). For a logarithmic potential in $d=3$
there are some subtleties.

Let us consider a problem with a potential like

\begin{equation}
\label{potential2}
V(x)=\frac{\Gamma\left(\frac{d-2-\eta_A}{2}\right)}{2^{\eta_A}(4
\pi)^{d/2}\Gamma\left(\frac{2+\eta_A}{2}\right)}\left
[\left(\frac{|x|}{a}\right)^{2-d+\eta_A}-1\right].
\end{equation}
Here we have taken into account
the effect of anomalous scaling due to matter-field fluctuations in our
original  problem.  A  logarithmic interaction corresponds to the case
$d=3$ and
$\eta_A = 1$, which is the case which eventually will be relevant for us.
Strictly speaking, the duality scenario in Section III is valid only at
$d=3$. However, as far as the scaling behavior is concerned, it is useful
to continue to the whole dimension interval $(2,4)$,
while keeping the same $\epsilon$-tensors. This dimensional continuation
procedure is reminiscent of the one considered in some RG studies of
Chern-Simons theories \cite{CFW}.  The recursion relations
we obtain are given by  (see Appendix B)

\begin{eqnarray}
\label{KNSrec}
&&~~~~~~~~~~~~~~~~~~~~~~\frac{dK^{-1}}{dl} =  y^2 -
(2-d+\eta_A)~ K^{-1} \nonumber \\
&&~~~~~~~~~~~~~~~~~~~~~~~~~~\frac{d y}{dl} =  \left[d-\eta_y-
2\pi^2\tilde{f}(d)K\right] ~ y,
\end{eqnarray}
where $\eta_y$ is the {\it anomalous dimension} of the fugacity which
is given by

\begin{equation}
\label{anomdimfug}
\eta_y=\frac{\eta_A}{2}=\frac{4-d}{2},
\end{equation}
and

\begin{equation}
\label{dimfactor}
\tilde{f}(d)=\frac{(d-2-\eta_A)\Gamma\left(\frac{d-2-\eta_A}{2}
\right)}{2^{\eta_A}(4\pi)^{d/2}\Gamma(1+\eta_A/2)}.
\end{equation}
Hence, for the case of a {\it logarithmic interaction in three dimensions},
which corresponds to $\eta_A=1$, the recursion
relations for the fugacity and the stiffness have  a similar structure
as the  standard Kosterlitz-Thouless recursion relations  one  obtains in
the two-dimensional case \cite{KT,Jose}.
The main difference is in the recursion relation for the
fugacity, which has an anomalous dimension $\eta_y=1/2$.
Note that the second term in the equation for $K^{-1}(l)$,
which prevents fixed points of the equations (\ref{KNSrec})
from being obtained, is absent for a  logarithmic potential in
any dimension.

When $\eta_A=0$, which corresponds to neglecting
the effect of matter fields in the original gauge theory, we
have $\eta_y=0$. Our recursion relations then
reduce to the ones given in Eqs. (\ref{KTrec1}) and (\ref{KTrec2})
obtained in \cite{Kosterlitz} by a very different method than
we employ in Appendix B.
Moreover, we have also derived Eqs. (\ref{KNSrec}) along
a different route than that used in Appendix B, namely
by the method employed in \cite{Jose}.
This constitutes an important consistency
check on our calculations.
For the case where $\eta_A=0$,
the absence of a phase transition reflects the permanent
confinement of electric test charges in the usual three-dimensional
compact QED \cite{Polyakov}.

We see that the flow equation for the fugacity obtained in
Eq. (\ref{KNSrec}) does not agree with the result of our
``spin-wave'' theory, which leads to Eq.(\ref{flowy1}) or,
equivalently, Eq. (\ref{flowy2}). {\it The reason for this is
that an anomalous scaling dimension $\eta_y$ for the fugacity is
induced by the renormalization of the stiffness}. Indeed, in
Appendix B we show that a potential like (\ref{potential1})
leads to an additional  scaling transformation  in the
{\it effective} stiffness of the form
$K(l)\to e^{(2-d+\eta_A)l} K(l)$.
If $\eta_A\neq 0$, this is compensated in the {\it effective}
fugacity by the scaling transformation,  $y(l)\to e^{-\eta_yl} y(l)$.
In the case of the Coulomb gas, where $\eta_A=0$, the spin-wave
analysis gives the right answer, Eq. (\ref{KTrec2}), as can easily
be seen by working out a Callan-Symanzik RG analysis in the sine-Gordon
theory (\ref{SG}) for arbitrary dimensions. Thus, deviations
from an ordinary type of Coulomb potential in $d$-dimensions
lead to an anomalous dimension to the fugacity, Eq. (\ref{anomdimfug}),
which cannot be obtained by spin-wave theory.

The important point to note here is that  a fixed point of the recursion
relations Eqs. (\ref{KNSrec}) for $d=3$ exists for the stiffness and
fugacity in
the limit of zero fugacity, so the problem scales to the weak coupling limit.
Hence, the problem is selfconsistently found to be amenable to a KT-type of
phenomenological RG analysis. It is not necessary to calculate to higher
order in $y$ to determine the fixed point. This  demonstrates that the
phase transition established above is of the KT type. This has some 
resemblance with the results of a rather
remarkable paper by Amit {\it et al.}
\cite{Amit}, which also finds a KT transition in a three-dimensional Coulomb
gas with logarithmic interaction between point charges (see Appendix A). 
In their case, the
logarithmic interaction between  the point charges in three dimensions
did not have its origin in anomalous scaling dynamically generated by
matter-field fluctuations, but originated in anisotropic higher order
derivative terms in an underlying field theory that were put in by hand. 
This anisotropy ultimately induces a dimensional reduction. 

In four dimensions, we have $\eta_A=0$ and extrapolating the above results
it is clear that no fixed points of the above recursion
relations can be found. Indeed, the above analysis no longer applies and
no KT topological phase-transition occurs. This is so because
by dualizing a compact Maxwell Lagrangian in four dimensions,
we obtain a non-compact Abelian Higgs model \cite{Peskin},
which cannot be brought onto the form of a Coulomb gas.
The transition in this case is known to be of more conventional
second or first order type \cite{FradShe}.

Finally, we note that in three dimensions there is a universal jump in the
stiffness parameter at the transition, analogous to what is known in the
$2d$ case \cite{NelKost}. In units of Eqs. (\ref{KNSrec}), this jump is
determined by  dimensionality and the anomalous scaling of the fugacity,
\begin{equation}
K_R\equiv\lim_{l\to\infty}K(l) = \frac{d-\eta_y}{2\pi^2\tilde{f}(d)}.
\label{Kjump}
\end{equation}

\section{RG functions of Abelian Higgs model and KT transition}

In this Section we show how the RG functions and fixed points in the
Abelian Higgs model are related to the KT-like transition described
in the previous Section. In particular, we shall use the critical coupling
$t_c$ to fix an {\it a priori} arbitrary constant that enters
into the computation of the critical exponents for the Abelian Higgs model.
This in our view improves on a scheme previously used \cite{Herbut}, where
a corresponding constant was fixed by appealing to numerical results for
the value of the Ginzburg-Landau parameter $\kappa$ which separates
first- from second-order behavior
\footnote{In an early Monte Carlo simulation, a tricritical value
$\kappa_{\rm{tri}}= 0.4/\sqrt{2}$ was found, \cite{Bartholomew}.
This is the value used in the {\it ad hoc} scheme
of Ref. \cite{Herbut}. More recently, a large-scale Monte Carlo simulation
improved on this value, finding $\kappa_{\rm{tri}}=(0.76 \pm 0.04)/\sqrt{2}$,
\cite{Mo1}.
This is in surprisingly good  agreement with an early analytical result
$\kappa_{\rm{tri}}=0.798/\sqrt{2}$, see Ref. \cite{Kleinert}. Using this
improved value for $\kappa_{\rm{tri}}$ in the $\beta$-functions of Ref.
\cite{Herbut}, the critical exponent $\nu$ obtained would be  $\nu = 0.53$.
This is quite far from the correct $3DXY$ value $\nu_{\rm{XY}}=0.67$, as
well as from the $3DXY$ one-loop value $\nu=0.625$.}. In our approach, the
parameter (denoted $r$ below) is fixed from our theory of the critical
behavior of the compact case, which we have argued in the Introduction
to be the same as for the non-compact Abelian Higgs model at infinite
{\it bare} gauge coupling.  Before doing this, however, a few preliminary
remarks are in order.

The Abelian Higgs model is manifestly a two-scale theory.
Indeed, the gauge field becomes massive due to the Higgs mechanism.
Thus, in the ordered phase we are left with two mass scales, the
Higgs mass $m$ and the gauge field mass $m_A$. From these two mass
scales we obtain the Ginzburg parameter $\kappa\equiv m/m_A$. Due to
the existence of two mass scales in the problem, we have very distinct
situations depending on whether $\kappa\ll 1$ or $\kappa\gg 1$.
For $\kappa\ll 1$ vortex lines, which are the topological defects
of the matter field, attract each other. This  corresponds
to a type I regime, while for $\kappa\gg 1$ we have repulsive
forces between vortex lines, which corresponds to the type II regime.
This two-scale behavior survives in the disordered phase, though
in this case $m_A=0$.

We shall consider the calculation of RG functions
for the massless theory, but using two renormalization scales \cite{Herbut}.
In order to see the influence of the two mass scales appearing in the
ordered phase, on the massless theory, we define the dimensionful couplings
at different renormalization points, $u$ at $\mu$ and $e^2$ at $\bar{\mu}$.
Let us define the ratio $r=\mu/\bar{\mu}$. By rewriting $e^2(\bar{\mu})$
in terms of $\mu$, we obtain the one-loop $\beta$-functions for any fixed
dimension $d\in(2,4]$ and an order parameter with $N/2$ complex components
\cite{KleinNog2}

\begin{equation}
\label{balphafixd}
\beta_\alpha=(4-d)[-\alpha+ r NA(d)\alpha^2],
\end{equation}

\begin{equation}
\label{bgfixd}
\beta_g=(4-d)\left\{-g+B(d)\left[-2(d-1)\alpha g+\frac{N+8}{2}g^2
+2(d-1)\alpha^2\right]\right\},
\end{equation}
where

\begin{equation}
A(d)=-\frac{\Gamma(1-d/2)\Gamma^2(d/2)}{(4\pi)^{d/2}\Gamma(d)},
\end{equation}

\begin{equation}
B(d)=\frac{\Gamma(2-d/2)\Gamma^2(d/2-1)}{(4\pi)^{d/2}\Gamma(d-2)}.
\end{equation}
From Eq. (\ref{balphafixd}) we see that
$\gamma_A=r(4-d)NA(d)\alpha$.
By considering $d=4-\epsilon$ and expanding for small $\epsilon$, we
recover the well known $\epsilon$-expansion result
\cite{HLM} if we take $r=1$. In our fixed dimension approach $r$
is an arbitrary parameter that is usually fixed by imposing additional
conditions \cite{Herbut}. When $d=3$ and $N=2$ we have the fixed point
$\alpha_*(r)=16/r$. In the context of the compact Abelian Higgs model
we fix the value of $r$ by demanding that $K_c$ should
correspond to a $r=r_c$, with
$K=1/\alpha_*$ at one-loop. If we use the spin-wave estimate
$K_c=3$ (which corresponds to $t_c=12\pi^2$),
we obtain then that $r_c=48$ and
thus $\alpha_*=1/3$. On the other hand, if we use the
estimate from our KT-like recursion relations, we have
$K_c=5/2$ and therefore $\alpha_*=2/5$.
In order to check the quality of these matchings,
we compute the critical exponents of the
three-dimensional Abelian Higgs model in $d=3$.
The critical exponent $\nu$ is given by the fixed point value of the
RG function

\begin{equation}
\label{nu}
\nu_\phi=\frac{1}{2+\gamma_m},
\end{equation}
where

\begin{equation}
\gamma_m=\mu\frac{\partial\ln Z_m}{\partial\mu}-\gamma_\phi,
\end{equation}
with $Z_m$ being the mass renormalization and

\begin{equation}
\gamma_\phi=\mu\frac{\partial\ln Z_\phi}{\partial\mu}.
\end{equation}
At the fixed point $\gamma_\phi$ gives the value of the critical
exponent $\eta$. At one-loop order, we have

\begin{equation}
\gamma_m=\frac{\alpha-g}{4}, ~~~~~\gamma_\phi=-\frac{\alpha}{4}.
\end{equation}
When $K_c=3$,
the fixed point for the coupling $g$ which corresponds to infrared
stability is given by $g_*=2(7+2\sqrt{11})/15$. Therefore, we obtain
$\nu\approx 0.615$ and $\eta=-1/12$. Using $K_c=5/2$, we obtain
$g_*=4(6+\sqrt{31})/25$. The critical exponents in this case
are $\nu\approx 0.61$ and $\eta=-1/10$.
Both estimates are close to the one-loop value
of the $XY$ model, $\nu_{XY}\approx 0.625$. From duality arguments
we expect indeed a $XY$ value for the exponent $\nu$ \cite{Kiometzis}.

\section{Summary and Discussion}

In this paper, we have considered the  Abelian Higgs model
in $2+1$ dimensions both for the non-compact and compact cases,
with matter fields in the fundamental representation.
We have performed a duality lattice transformation on these
models, emphasizing the features that set them apart as
well as those they have in common. A major difference lies
in the fact that in the dual formulation, the non-compact case
has stringent constraints ${\bf \nabla} \cdot {\bf l}_i = 0$
imposed on the topological currents of the system, while in
the  compact case ${\bf \nabla} \cdot {\bf l}_i$ can take
any integer value, i.e. the currents are  unconstrained
{\it for the case where the matter field is in the fundamental
representation}. This effectively makes the dual non-compact
case a much more strongly interacting system of topological
currents, and this is why phase transitions are more easily
brought out compared to the compact case. As a result, we have seen that there
is one limit of the LAH model where the non-compact case exhibits the
$IXY$ transition, while the compact case is  an exactly
soluble discrete Gaussian model with apparently no phase transition.

A major part of the paper (Section III and IV) has been devoted to
establishing that, despite the absence of any phase transitions with
a local order parameter
in the compact case, a topological phase transition nevertheless
is found in the interior of the phase diagram of the model.
A key ingredient is the renormalization of the gauge-field
propagator of the problem due to critical matter field
fluctuations, Eq. (\ref{LA}). With no matter fields present, the
topological defects of the gauge field, which are monopole
configurations, interact with a $1/R$ potential in $d=3$. In the
presence of matter fields, taking into account their critical fluctuations,
the resultant effective gauge theory may be described as
an overall neutral plasma of charges that interact with
a logarithmic potential in $d=3$, Eq. (\ref{monopaction}).
A field-theoretical formulation of  the action given
in Eq. (\ref{monopaction}) yields an {\it anomalous} sine-Gordon
(ASG) model, Eq. (\ref{CPSG}).
A  renormalization group analysis of  this model
based on the Callan-Symanzik equations
shows that the theory is massive below a critical value of the
coupling constant. This by itself suffices to conclude that a phase
transition exists. We then go on to show that the problem
is amenable to an analysis based on KT-like recursion relations, Eqs.
(\ref{KNSrec}),  derived for a $d$-dimensional
gas of point charges interacting with a pair-potential which in a certain limit is
logarithmic. In this limit, the recursion relations we derive
for the stiffness and fugacity of the problem  reduce
to equations which are similar in structure to the well known
Kosterlitz-Thouless recursion relations obtained for the  two-dimensional Coulomb
gas, but with a modified equation for the fugacity
due to an induced anomalous scaling of it. This anomalous scaling
in the fugacity accounts for  deviations from the ordinary Coulomb gas
case in $d$ dimensions.
The change in the equation for the fugacity shows
that  the stiffness and the fugacity
of the problem  mutually influence each other under renormalization
in a manner which
is different from the case of a logarithmic pair-interaction in $d=2$.
As a consequence of this, the universal jump in the stiffness at
the transition is then  given, in appropriate units, by the
dimensionality of the system and the anomalous scaling of the
fugacity, Eq. (\ref{Kjump}).

In Section V, we have seen that the deconfinement phase transition we
find in the compact case, with a critical coupling $t_c$,
allows us to fix a parameter appearing in the evaluation of the critical
exponents of the non-compact Abelian Higgs model.
This represents an improvement on previous schemes to fix this
parameter.

We close with a few remarks on unsolved problems.  When
only fermionic fields are coupled to the massless gauge field
(spinor $QED_3$), then we again
obtain a  $\beta$ function for the renormalized gauge coupling as
given in Eq. (\ref{betaalpha}), but $\gamma_A$ in the equation now
depends only on one coupling constant, $\alpha$, not two as in the bosonic
case. Then we do not have the freedom to tune parameters of the model to
drive it through a phase transition of the type described in Section IV.
The analysis of Section IV may be carried through as before, but the point
is  that the fixed point coupling, $\alpha = \alpha_*$ does not depend
on any second coupling constant $g$, this simply does not appear in the theory.
Instead, $\alpha_*$ depends on the number of fermion flavours
$N$ only. In principle there thus exists a critical value  $N=N_c$
where  the compact version of the model with fermionic matter, also
goes through a deconfinement transition. The confining phase
corresponds to $N<N_c$. It is highly controversial what
this critical value is. A simple one-loop renormalization group calculation
gives $N_c=24$ \cite{KNS} in agreement with an earlier result by Ioffe and
Larkin obtained by a quite different method \cite{IL}. However, we may in
fact expect that the actual value is much smaller than this. Marston has
calculated the same number using one-instanton action and finds $N_c=0.9$
\cite{Marston}. The important point here is that whatever the precise
value of $N_c$ is, the interaction between the monopoles is
always logarithmic.

Also, in the fermionic case there is a subtlety in that another
type of instability, absent in the bosonic case, could intervene to
destroy the  deconfinement transition. Fermions can in principle undergo
a spontaneous chiral symmetry breaking (S$\chi$SB) \cite{Pisarski}.
This happens when the number of fermion flavours
is less than some critical value, $N_{{\rm ch}}$ say.  This means that
a fermion mass is dynamically generated for $N<N_{{\rm ch}}$. The precise
 value of $N_{{\rm ch}}$ is presently also a matter of debate.
One estimate from the Schwinger-Dyson equation gives $N_{\rm ch}=32/\pi^2$
\cite{Appelquist1}. This result is confirmed by Monte Carlo simulations
finding $N_{\rm ch}\approx 3.5$ \cite{DKK}. Another analytic calculation
gives $N_{\rm ch}=128/3\pi^2$ \cite{Appelquist2}. A recent estimate
based on a new constraint on strongly interacting systems
gives  $N_{\rm ch}\leq 3/2$ \cite{Appelquist3}.
This is quite consistent with the most recent numerical results we
are aware of \cite{Hands}, where no signs of S$\chi$SB is
found for $N \geq 2$.
Thus, there is
no consensus on the precise value of $N_{\rm ch}$. The calculation
of $N_c$ assumes that the fermions are massless. Thus, if
$N_c=24$ as in Refs. \cite{KNS} and \cite{IL}, then a
deconfinement transition will take place since
the fermion mass is generated at a much lower value of $N$.
With massive fermions present our anomalous three-dimensional compact
QED scenario does not apply because the Maxwell term does not
become irrelevant anymore. In such a situation the results of Polyakov
\cite{Polyakov} apply and there is permanent confinement of electric test
charges. This would be the case for the value $N_c=0.9$ obtained by
Marston \cite{Marston}, which lies below all estimates of
$N_{\rm ch}$. In this case the deconfinement transition does
not happen.

Physically, S$\chi$SB in spinor $QED_3$ has important
consequences in the physics of high-$T_c$ cuprates. As we mentioned
in the Introduction, spinor $QED_3$ with a compact gauge field emerges
as a possible low energy description of the fluctuations around the
flux phase in the quantum Heisenberg antiferromagnet \cite{Affleck}.
In this context, the dynamical mass generation is associated with the
spin density wave (SDW) instability. Thus, gauge field fluctuations
could in principle restore the N\'eel state. The physical number of
fermion components in this case is $N=2$. Spinor $QED_3$ also emerges
by considering the low energy physics of the $d$-wave superconducting
state in the pseudogap phase of the  high-$T_c$ cuprates \cite{Tesanovic}.
In this case, however, the gauge field is non-compact and there
is an inherent anisotropy in the Lagrangian. There also, S$\chi$SB is
responsible for the onset of SDW as half-filling is approached \cite{Herbut1}.
The physical number of fermion components in this case is again
$N=2$. Therefore, in these theories it is essential that
$N_{\rm ch}>2$. If the most recent estimate for $N_{\rm ch}$ is
correct \cite{Appelquist3}, this could have serious implications
for the validity of the different spinor $QED_3$ scenarios
discussed above. In the case of the spinor $QED_3$ description of
the pseudogap phase, the inherent anisotropy could possibly affect
the value of $N_{\rm ch}$. However, results presented
thus far indicate that at least weak anisotropy  will
not affect $N_{\rm ch}$ obtained in the isotropic case
\cite{Herbut2}.    Moreover,
when studying effective theories of  undoped high-$T_c$
cuprates, we have argued in the Introduction that the
relevant theory to study is fermions coupled to
compact $U(1)$ gauge-fields. Hence, it is of importance
to revisit the problem of how  monopoles affects S$\chi$SB
\cite{Fiebig}. Finally, we note that a recent provocative
paper by Wen \cite{Wen2} states that there exists a
principle of {\it quantum order} which may prevent
fermions from dynamically acquiring a mass even in the
presence of strong coupling to gauge fields. Hence, it seems
to us that a renewed effort in numerical computations of
$N_{\rm ch}$ in $2+1$-dimensional gauge theories coupled
to fermionic matter, including  the effects of compactness
and anisotropy, would be very timely.

\section*{Acknowledgments}
The authors thank M. Chernodub,
P. C. Hemmer, J. S. H{\o}ye, J. M. Kosterlitz,
J. B. Marston, and D. R. Nelson  for useful communications, and in
particular S. Sachdev for stimulating remarks. F. S. N. and A. S. thank
O. Sylju{\aa}sen and NORDITA, where part of this work has been done,
for hospitality. A. S. would also like to thank the
Institute for Theoretical Physics at the Free University Berlin
for the hospitality in the early stages of this work.
The work of F.S.N. is supported by the Alexander von
Humboldt foundation. A.S. acknowledges support from the Norwegian
Research Council through grant No. 148825/432 and from 
the ESF program {\it Vortex Physics at Extreme Scales and 
Conditions}. The research group of H.K. receives 
funds from ESF under the program {\em Cosmology in the Laboratory}. 

\appendix

\section{KT-like transition in three dimensions in an anisotropic
sine-Gordon theory}

While considering a class of globally symmetric self-dual
$Z_N$ models in the $N\to\infty$ limit,
Amit {\it et al.} \cite{Amit} arrived at the following
anisotropic three-dimensional
sine-Gordon action containing higher derivatives:

\begin{equation}
\label{aniSG}
S_{\rm ANISG}=\int d^3 x\left[\frac{1}{2t_1}(\partial_\parallel^2
\varphi)^2+\frac{1}{2t_2}(\partial_z\varphi)^2-z\cos\varphi\right],
\end{equation}
where $\partial_\parallel^2=\partial_x^2+\partial_y^2$. As
pointed out in Ref. \cite{Amit}, the above model has a
KT transition in three dimensions. Indeed, it is easy to see
that the propagator is logarithmic at large distances.
Note, however, that anisotropy and the higher order derivatives
in the parallel direction are essential, and the system
effectively shows two-dimensional behavior by dimensional 
reduction. This is in contrast
with our genuinely three-dimensional KT-like scenario.

\section{KT-like  recursion relations}

In this appendix we derive to lowest order in the fugacity the recursion
relations for  the scale-dependent stiffness parameter $K(l)$ and
fugacity $y(l)$ given in Eqs. \ref{KNSrec} for a $d$-dimensional plasma
where the bare pair-potential is given by Eq. (\ref{potential2}),
which reduces to a logarithmic potential when $d=3$.
The starting point will be a low-density approximation for
a dielectric constant of this system. We closely follow
a method for doing this introduced in \cite{Young}.
Introducing the solid angle  in $d$ dimensions
$\Omega_d = 2 \pi^{d/2}/\Gamma(d/2)$ and the density of dipoles
in the fluid by $n_d$,  a low-density approximation
for the dielectric constant is given by
\begin{equation}
\varepsilon = 1 + n_d ~\Omega_d  ~\alpha
\end{equation}
where $\alpha$ here denotes the polarizability of the medium,
a standard linear-response analysis gives
$\alpha = 4\pi^2 K <s^2> / d$
and $<s^2>$ is the mean square of the dipole moment
in the system. To compute this, we need the
low-density limit of the pair-distribution function
$n^\pm (r)$ of the   plasma, which is readily obtained from
the grand canonical partition function $\Xi$ expanded
to second order in the bare fugacity $\zeta$, and replacing
the thermal de Broglie wavelength by a short-distance cutoff
$r_0$, as  follows
\begin{equation}
n^\pm (r) = \frac{\zeta^2}{r_0^{2d}} ~e^{-4\pi^2K V}.
\end{equation}
In this way, we may now go on to express a {\it scale-dependent} dielectric
constant as follows
\begin{equation}
\varepsilon(r) = 1 + \frac{4\pi^2\Omega_d K}{d} \int_{r_0}^r ~ds ~ s^{d+1} ~n^\pm (s).
\label{dielscal}
\end{equation}
Note however, that in Eq. (\ref{dielscal}), a mean-field approximation is
understood to be used by replacing the bare potential $V$  in $n^\pm (r)$ by an
{\it effective
potential $U(r)$}. This effective screened potential must be selfconsistently
determined by demanding that it  gives rise to an electric field in the problem
given by
\begin{equation}
\frac{\partial U}{\partial r} = E(r) = \frac{\tilde{f}(d)}
{\varepsilon(r)~ r^{1-\rho}},
\label{field}
\end{equation}
where $\rho=2-d+\eta_A$ and $\tilde{f}(d)$ is defined in Eq.
(\ref{dimfactor}).
Such a mean-field procedure has been consistently used with success
in the $2d$ case, and the origin of the success lies in the long range
of the $\ln$-interaction. In higher dimensions, such a
procedure will work even better since the logarithmic
potential is felt over even longer distances due to extra
volume factors.

Let us introduce a logarithmic length scale $l=\ln(r/r_0)$ along with
the new variables
\begin{eqnarray}
~~~~~~~~~~~~~~~~~~~~\tau(l) & = &
\frac{ \varepsilon(r_0\exp l)}{4\pi^2K} \nonumber \\
~~~~~~~~~~~~~~~~~~~~x(l) & = &  {4\pi^2K} U(r_0\exp l)
\end{eqnarray}
Here, $x(l)$ is determined
selfconsistently by integrating the effective field $E(r)$.
Then we get from Eqs. (\ref{dielscal}) and (\ref{field})
\begin{equation}
\tau(l)=\tau(0) + \frac{\Omega_d ~ \zeta^2}{d r_0^{d-2}}
\int_{0}^l ~dv ~~ e^{(d+2) v - x(v)},
\label{tau}
\end{equation}
and
\begin{equation}
x(l) = x(0) + \tilde{f}(d)
\int_0^l ~dv ~ \frac{r_0^\rho ~ e^{ \rho v}} {\tau(v)}.
\label{x}
\end{equation}
From Eqs. (\ref{tau}) and (\ref{x}), we may derive coupled renormalization
group  equations for $\tau(l)$ and $x(l)$. However, in order to obtain
equations that have a form more similar to equations that have  appeared in
the literature on the $d$-dimensional Coulomb gas \cite{Kosterlitz}, we
introduce a new variable $K(l)$  representing a scale dependent
stiffness constant, as follows
\begin{equation}
K^{-1}(l)  \equiv  \frac{\tau(l)}{r_0^{\rho} ~ e^{\rho l}}.
\label{kl}
\end{equation}
Thus, we see that the effect of a nonzero $\rho$ on the
stiffness amounts to a scaling change $K(l)\to e^{\rho l} K(l)$.
Using Eq. (\ref{x}), we have that
\begin{equation}
\label{rel}
\frac{\partial x(l)}{\partial l} = 4\pi^2\tilde{f}(d) K(l).
\end{equation}
Differentiating $K^{-1}(l)$  with respect to $l$ and using Eq. (\ref{tau}), we obtain
\begin{equation}
\label{knsflowk}
\frac{\partial K^{-1}(l)}{\partial l} = - \rho ~ K^{-1}(l)
+ \frac{2\Omega_d\zeta^2}{d r_0^{d-2+\rho}} ~e^{[(d+2 -\rho)l - x(l)]}.
\end{equation}
From this expression, we define a scale dependent fugacity $y(l)$ given by
\begin{equation}
\label{fugty}
y(l)  \equiv   \frac{\sqrt{2
\Omega_d} ~\zeta ~ e^{[(d+2-\rho)l - x(l)]/2}}{
\sqrt{d}r_0^{(d-2 +\rho)/2}}.
\end{equation}
Thus, we see explicitly that the renormalization of $K(l)$ in principle influences
the flow equation for $y(l)$, which is obtained by differentiating with
respect to $l$ and using Eq. (\ref{rel})
\begin{equation}
\label{knsflowy}
\frac{\partial y(l)}{\partial l} =
\left[d-\eta_y - 2\pi^2\tilde{f}(d)K(l) \right] ~ y(l),
\end{equation}
where $\eta_y=(d-2+\rho)/2$.
Eqs. (\ref{knsflowk}) and (\ref{knsflowy}) are precisely Eqs. (\ref{KNSrec}).
On the other hand, the Callan-Symanzik approach of Section IV-A, which
basically ignores the influence of {\it the renormalization of $K(l)$}
on the structure of the flow equation for $y(l)$,  yields as we have seen
Eq. (\ref{flowy2}). We have already remarked in Section IV-B that
this type of approach gives the correct answer only if there are no
deviations from the Coulomb potential case, that is, we need
$\rho=2-d$. Note that in the usual KT transition we would have
$\rho=\eta_y=0$.

\section{Screened effective potential }
In this appendix, we derive the asymptotic long-distance behavior of
the screened effective interaction $U(r)$ introduced in Appendix B,
for the case $\rho =0$,
corresponding to $d=3$ and
$\eta_A=1$. We start from the recursion relations, written
on  the form
\begin{eqnarray}
~~~~~~~~~
~~~~~~~~~
\frac{\partial K^{-1}}{\partial l }
=  y^2 ,
~~~~\frac{\partial y}{\partial l}
= \left[\frac{5}{2} - K (l)\right] y.
\end{eqnarray}
From Eq. (\ref{kl}),
we have that $K^{-1}(l) = \tau(l)$ in this case. Next, we introduce
the variable $T(l)$ defined by
\begin{eqnarray}
~~~~~~~~~~~~~~~~~~~T(l) & \equiv & \frac{5
~  \tau(l)/2 - 1 }{
5~ \tau(l)/2} \nonumber \\
~~~~~~~~~~~~~~~~~~~& \approx &\frac{5}{2}  ~ \tau(l) - 1,
\label{Tl}
\end{eqnarray}
where the latter approximation is asymptotically exact close enough
to the transition. In terms of this, the flow equation for the fugacity
may be written on  the form
\begin{equation}
\frac{\partial y^2(l)}{\partial l} =  ~ 5 ~ T(l)~ y^2(l),
\end{equation}
On the other hand, we have
\begin{eqnarray}
~~~~~~~~~~~~~~~~~~~~\frac{\partial T^2 (l) }{\partial l}
& \approx &  5 ~T(l) ~ \frac{\partial \tau(l)}{\partial l} \nonumber \\
~~~~~~~~~~~~~~~~~~~~& = &  ~ 5 ~ T(l) ~ y^2 (l),
\end{eqnarray}
and hence we have
\begin{equation}
y^2(l)-T^2 (l) = \pm \omega^2,
\label{firstint}
\end{equation}
where $\omega$ is some positive number. We are interested in the quantity
$\lim_{l \to \infty} x(l)$ for the case where $y^2(l)-T^2(l)<0$, and
$T(l) < 0$, this will be the regime where the fugacity scales to zero.
In this case we choose the negative sign on the r.h.s. in Eq. (\ref{firstint}).
From the flow equation for $K^{-1}(l)$ we find

\begin{eqnarray}
~~~~~~~~~~\frac{\partial T(l)}{\partial l}
& = & \frac{5}{2} ~ y^2(l)
 = -\frac{5}{2} ~ \left[ \omega^2 - T^2 (l) \right].
\end{eqnarray}
This is solved to obtain, introducing $u=(5/2)~  \omega ~ l + \theta$,
\begin{eqnarray}
~~~~~~~~~~~~~~~~~~T(l) & = & -\omega ~\coth u \nonumber \\
~~~~~~~~~~~~~~~~~~y(l) & = & \frac{\omega}{\sinh u},
\label{solutions}
\end{eqnarray}
where $\omega$ and $\theta$ are integration constants that are uniquely
determined from the initial conditions on $\tau(l)$ and $y(l)$, i.e.
by the bare coupling constants of the problem as follows
\begin{eqnarray}
~~~~~~~~~~~~~~~~y^2(0) - T^2 (0) & = & -\omega^2 \nonumber \\
~~~~~~~~~~~~~~~~\frac{T(0)}{y(0)} & = & - \cosh\theta.
\label{initcond}
\end{eqnarray}
From the expression for $T(l)$,  using Eq. (\ref{Tl}), we obtain
\begin{equation}
\tau(l) = \frac{2}{5}(1 - \omega ~ \coth u).
\end{equation}

Since $\tau(l) > 0$, this puts restrictions on the constants
$\omega$ and $\theta$, and the most severe limitations on
$\omega $ in terms of $\theta$ is given by
\begin{equation}
1 - \omega \coth\theta > 0.
\label{condition}
\end{equation}
Using Eq. (\ref{rel}) and the fact that $K(l) = 1/\tau(l)$, we have
\begin{equation}
\frac{\partial x(l)}{\partial l} =
\frac{5/2}{1-\omega ~ \coth u}
\label{derxl}
\end{equation}
From Eq. (\ref{condition}), we see  that $\partial x(l)/\partial l > 0$.
This is an important result, since it immediately reveals that,
in the regime $y^2(l)-T^2 (l) < 0$ we consider here, the logarithmic
bare potential $V(r)$ cannot possibly be screened into a power law potential
$1/r^\sigma$ with $\sigma >0$, since in that case we would have
$\partial x(l)/\partial l <0$. However, for all $l$ we have
\begin{equation}
\frac{\partial^2 x(l)}{\partial l^2} = -
\left(\frac{5 ~ \omega/2}{ \sinh u - \omega ~ \cosh u } \right)^2 < 0.
\end{equation}
Introducing $\omega_{\pm} = 1 \pm \omega$, Eq. (\ref{derxl}) is straightforwardly
integrated to yield
\begin{equation}
x(l)- x(0) = \frac{1}{\omega_+\omega_-}\left[\frac{5}{2}\omega_+ l
+\ln\left(\frac{\omega_+e^{-2\theta}+\omega_-}{\omega_+e^{-2u}+\omega_-}
\right)\right].
\end{equation}
From this, it follows that for $r\gg r_0$ the effective potential 
behaves asymptotically as 
\begin{equation}
U(r) \sim \ln(r/r_0).
\end{equation}

\section{Exact equation of state for the $d$-dimensional $\ln$-plasma}

The equation of state for a $d$-dimensional $\ln$-plasma with no
short-distance cutoff, may be obtained via a simple scaling argument,
previously applied to the two-dimensional case \cite{Hauge}. The configurational
integral in the canonical partition function is given by
\begin{equation}
Q = \int_V \cdots\int_V d^d {\bf r}_1 \cdots d^d {\bf r}_{2N}
\exp[{\tilde t} \sum_{i < j} q_i q_j \ln(r_{ij})]
\end{equation}
where $q_i = \pm 1$, and we assumed that we have $2N$ particles in the
system, $N$ with charge $q_i=1$ and $N$ with charge $q_i=-1$,
$\sum_{i=1}^{2N} q_i = 0$. Here,  $V = L^d$ is the volume of the system.
Introduce new dimensionless variables $R_{ij} = r_{ij}/L$
where $r_{ij} = |{\bf r}_i - {\bf r}_j|$, in which
case the configurational integral is given by
\begin{eqnarray}
Q &   =  & L^{2 N d}  \int_0^1 \cdots\int_0^1  ~ d^dR_1 \cdots d^dR_{2N}
\exp({\tilde t} \sum_{i < j} q_i q_j ~ \ln(R_{ij} L)) \nonumber \\
  &   =  & L^{2 N d}
\exp[{\tilde t} \sum_{i < j} q_i q_j  \ln(L)  ] ~I,
\end{eqnarray}
where the integral $I$ is independent of volume.
Now note that
\begin{eqnarray}
2 \sum_{i < j} q_i q_j & = & \sum_{i \neq j} q_i q_j
 =  (\sum_i q_i) ( \sum_j q_j) - \sum_{i=1}^{2N} q_i^2 \nonumber  \\
& = &-2 N
\end{eqnarray}
Then we obtain
\begin{eqnarray}
~~~~~~~~~~~~~~~~Q & = & L^{2 N d} e^{- {\tilde t} N \ln(L)}~I
= L^{2Nd -{\tilde t} N} ~I  \nonumber \\
~~~~~~~~~~~~~~~~& = & V^{2N - {\tilde t} N/d} ~ I.
\end{eqnarray}
From this,  we obtain the equation of state involving the pressure
\begin{equation}
{\tilde t}~ p ~ V = 2N - \frac{ {\tilde t} ~  N}{d}.
\label{eqstate}
\end{equation}
Note that the pressure vanishes when ${\tilde t} = {\tilde t}_0 = 2 d$.
A prerequisite for the validity of the above analysis is
that the quantity $I$ must be finite, otherwise the scaling
of variables that lead to the equation of state is
meaningless. In fact, $I$ is not always finite. Consider
again the integrand in $Q$, which is given by a product
of factors
\begin{equation}
e^{{\tilde t} \sum_{i<j} q_i q_j ~\ln(r_{ij})}
= \prod_{i < j} r_{ij}^{{\tilde t} q_i q_j}
\end{equation}
Any factor with $q_i = -q_j$ will be singular
when $r_{ij}=0$, which is possible in the
absence of a short-distance cutoff. To investigate whether
or not this singularity is integrable, consider the integral
\begin{equation}
\int dr r^{d-1} ~ r^{-{\tilde t}}
\end{equation}
This is finite only if
\begin{equation}
d - {\tilde t} > 0
\end{equation}
This means that the equation of state Eq. (\ref{eqstate}) makes sense
for ${\tilde t} < {\tilde t}_c = d$, note that for all
dimensions  $d$, ${\tilde t}_0 = 2 {\tilde t}_c$.

In two dimensions, it is known that the negativity of the
pressure occurs at a temperature that coincides with the
KT vortex-antivortex unbinding temperature, and that there
is a phase transition at twice this temperature. It is amusing
to note here that in the three-dimensional case,
the pressure vanishes at $t_c = 12 \pi^2$, after
having reintroduced ${\tilde t} = t/4\pi^2$. This is precisely
the critical coupling we found in Section IVA from
the Callan-Symanzik equations. In addition
there is again a phase transition at precisely
half the value of
this coupling constant, where the pressure becomes that of
an ideal gas of $N$ particles. In arbitrary dimensions,
this persists, the phase transition to an ideal gas
of $N$ particles always happens at half of the value
at which the pressure vanishes. This  phase transition,
which is a collapse of an overall charge-neutral
plasma of $N$ $q_i = +1$ charges and $N$ $q_i = -1$ charges into
an ideal gas of $N$ particles,
occurs because of the lack of a short-distance cutoff
in the system we consider in this appendix.

\section{Duality in the Abelian compact Higgs model with a
Chern-Simons term}

For completeness,
we present
in this Appendix the duality transformation
of the LAH with a Chern-Simons term added  \cite{Jackiw}. Compact
gauge theories with Chern-Simons term added are relevant in studies
of chiral spin liquid states \cite{Wen} when spinor states have been
integrated out. Such theories have been argued to exhibit a deconfinement
transition \cite{Schaposnik,Diamantini}. The  compact LAH  mode, i.e.
$A_{i\mu}\in(-\pi,\pi)$, with a Chern-Simons term has the action

\begin{eqnarray}
\label{LCSAH}
S_{\rm CS}&=&\sum_i\left[\frac{\beta}{2}(\nabla_\mu\theta_i-A_{i\mu}-
2\pi n_{i\mu})^2
+\frac{1}{2e^2}(\epsilon_{\mu\nu\lambda}
\nabla_\nu A_{i\lambda}-2\pi N_{i\mu})^2
\right.\nonumber\\
&+&\left.i\frac{\gamma}{2}(\nabla_\mu\theta_i-A_{i\mu}-
2\pi n_{i\mu})(\epsilon_{\mu\nu\lambda}
\nabla_\nu A_{i\lambda}-2\pi N_{i\mu})\right].
\end{eqnarray}
Let us introduce auxiliary fields ${\bf a}_i$, ${\bf b}_i$,
$\lambda_{i\mu}$, and $\sigma_{i\mu}$, such that

\begin{eqnarray}
S_{\rm CS}'&=&\sum_i\left[\frac{\beta}{2}{\bf a}_i^2
+\frac{1}{2e^2}{\bf b}_i^2+i\frac{\gamma}{2}
{\bf a}_i\cdot{\bf b}_i
+i\lambda_{i\mu}(\nabla_\mu\theta_i-A_{i\mu}-2\pi n_{i\mu}-a_{i\mu})
\right.\nonumber\\
&+&\left.i\sigma_{i\mu}(\epsilon_{\mu\nu\lambda}
\nabla_\nu A_{i\lambda}-2\pi N_{i\mu}-b_{i\mu})\right].
\end{eqnarray}
Next we introduce integer valued fields $m_{i\mu}$ and
$M_{i\mu}$ via the Poisson formula:

\begin{eqnarray}
S_{\rm CS}''&=&\sum_i\left[\frac{\beta}{2}{\bf a}_i^2
+\frac{1}{2e^2}{\bf b}_i^2+i\frac{\gamma}{2}
{\bf a}_i\cdot{\bf b}_i
+i m_{i\mu}(\nabla_\mu\theta_i-A_{i\mu}-a_{i\mu})
\right.\nonumber\\
&+&\left. i M_{i\mu}
(\epsilon_{\mu\nu\lambda}
\nabla_\nu A_{i\lambda}-b_{i\mu})\right].
\end{eqnarray}
Integration of $\theta_i$ and $A_{i\mu}$ give the constraints
enforced by delta of Kronecker

\begin{equation}
\nablab\cdot{\bf m}_i=0,
\end{equation}

\begin{equation}
\nablab\times{\bf M}_i={\bf m}_i.
\end{equation}
Summing over ${\bf m}_i$ gives

\begin{equation}
S_{\rm CS}'''=\sum_i\left[\frac{\beta}{2}{\bf a}_i^2
+\frac{1}{2e^2}{\bf b}_i^2+i\frac{\gamma}{2}
{\bf a}_i\cdot{\bf b}_i
-i(\nablab\times{\bf M}_i)\cdot{\bf a}_i
-i{\bf M}_i\cdot{\bf b}_i\right]
\end{equation}
By integrating out ${\bf a}_i$ and ${\bf b}_i$ we arrive
at the action

\begin{equation}
\tilde{S}_{\rm CS}=\frac{K}{2}\sum_i\left[
(\nablab\times{\bf M}_i)^2+\beta e^2{\bf M}_i^2
-ie^2\gamma{\bf M}_i\cdot(\nablab\times{\bf M}_i)
\right],
\end{equation}
where $K\equiv 4/(\gamma^2e^2+4\beta)$.
Using the Poisson formula to introduce a real lattice field $h_{i\mu}$
and doing an appropriate rescaling of the variables we obtain finally
the partition function

\begin{equation}
\label{chern-simons}
Z=Z_0\sum_{\{{\bf l}_i\}}\int_{-\infty}^{\infty}
\left[\prod_{i,\mu}dh_{i\mu}\right]\exp[-S_{\rm CS}^{\rm dual}({\bf h}_i,
{\bf l}_i)],
\end{equation}
where

\begin{equation}
\label{CSdual}
S_{\rm CS}^{\rm dual}=
\frac{K}{2}\sum_i\left[(\nablab\times{\bf h}_i)^2
+\beta e^2{\bf h}^2
-i\gamma e^2{\bf h}_i
\cdot(\nablab\times{\bf h}_i)\right]
+i2\pi {\bf l}_i\cdot{\bf h}_i.
\end{equation}
which should be compared with Eqs. (\ref{Z4}) and (\ref{compZ2}).
Note the appearance of the cross-term $i\gamma e^2{\bf h}_i
\cdot(\nablab\times{\bf h}_i)$. When the ${\bf h}_i$ are integrated out we
are thus left with a partition of the same form as Eq. (\ref{compdual1}),
but with an asymmetric propagator.

If we were to consider the non-compact LAH with a Chern-Simons term
added, and in the absence of the Maxwell term, $e^2 \to \infty$,  then
this is an effective  description of the fractional quantum Hall effect
\cite{Fradkin_book,ReyZee}. In this case we obtain
\begin{equation}
\label{CSdual1}
S_{\rm CS}^{\rm dual}=
\sum_i\left[\frac{1}{2 \beta} (\nablab\times{\bf h}_i)^2
-\frac{i}{2 \gamma} {\bf h}_i
\cdot(\nablab\times{\bf h}_i)\right]
+i2\pi {\bf l}_i\cdot{\bf h}_i.
\end{equation}
This is essentially the same as Eqs. (\ref{chern-simons}) and (\ref{CSdual})
for the compact case (with no mass term for the  ${\bf h}_i$-fields), but
we should add an additional constraint in the ${\bf \nabla} \cdot {\bf l}_i=0$
in the partition function.

One point worth emphasizing here, sometimes overlooked, is that the
gauge-field ${\bf h}_i$ is never a compact gauge-field, whether one starts
from an original compact or non-compact gauge theory. In the non-compact
Chern-Simons theory,  there exists a self-dual point at a value
$\gamma=1/2 \pi$ \cite{ReyZee,Nogueira1}. The possibility of self-duality
is a consequence of non-compactness, it can never arise starting from a
compact LAH model with Chern-Simons term added. It is an intriguing question
whether the self-duality at the above particular value of $\gamma$ in the
non-compact case corresponds to a critical point. A candidate physical
interpretation of such a putative phase transition would correspond to
statistical transmutation of the Laughlin quasiparticles of the
fractional quantum Hall effect as magnetic field is varied, since
in the context of the FQHE, the parameter $\gamma$ depends on filling
fraction, i.e. magnetic field. It is known that for the half-filled
lowest Landau level, the quasiparticles are fermions \cite{HLR}, while for
other filling fractions they are anyons.

\end{document}